\documentclass[11pt]{article}
\usepackage[top=1in, bottom=0.7in,left=1in, right=1in]{geometry}
\usepackage{amssymb,amsmath}
\usepackage{times}
\usepackage{amsthm}
\usepackage[dvips]{graphicx}
\usepackage{subfigure}
\usepackage{float}
\usepackage{setspace}
\usepackage[centerlast,small,bf]{caption}
\newtheorem{thm}{Theorem}[section]
\newtheorem{lem}[thm]{Lemma}

\begin{document}
\date{}

\title{A simple linear-time algorithm for generating auxiliary 3-edge-connected subgraphs}

\author{
Yung H. Tsin\footnote{School of Computer Science,  University of Windsor, Windsor, Ontario, Canada, N9B 3P4; peter@uwindsor.ca.
}\\
}

\maketitle \thispagestyle{empty}

\begin{abstract}

A linear-time algorithm for generating auxiliary subgraphs corresponding to the 3-edge-connected components of a connected multigraph is presented. The algorithm employs an innovative graph contraction operation combined with depth-first search, requiring only a single pass over the graph. In contrast, the best-known previous algorithms necessitate multiple passes: first to decompose the graph into its 2-edge-connected or 2-vertex-connected components, then to further break down these components into their 3-edge-connected or 3-vertex-connected components, and finally to construct a cactus representation of the 2-cuts to derive the auxiliary subgraphs for the 3-edge-connected components.

\vspace{30pt}

\end{abstract}

\doublespacing

\newpage

\noindent

\section{Introduction}

Given an undirected connected graph $G = (V,E)$.
 An \emph{edge-cut} of $G$ is a set of edges $S \subseteq E$ whose removal results in a disconnected graph.
 The edge-cut $S$ is a $k$-\emph{cut} if its cardinality $|S|= k$.
A $k$-\emph{edge-connected graph} is a graph that   has no $h$-\emph{cut}, for $h < k$.
 An edge-cut $S$ \emph{separates two vertices u and v}, if $u$ and $v$ lie in different
connected components after $S$ is removed from $G$.
Two vertices \emph{u and v are k-edge-connected} in $G$, denoted by $u \overset{k}{\equiv}_G v$,
if there are $k$-edge-disjoint paths connecting them.
A $k$-\emph{edge-connected component} (abbreviated as \emph{kecc}) of
$G$ is a maximal set of vertices $C \subseteq V$ such that
every two vertices in it are connected by  $k$ edge-disjoint paths in $G$, or equivalently,
there is no $(k-1)$-cut that separates two vertices in the set.

For $k \le 3$, linear-time algorithms for computing $k$eccs have been known for quite some time.
On the other hand,
the problem of computing the 4eccs of an undirected graph in linear time remained open for over two decades until recently.
Nadara et al.~\cite{NRSS21} presented a linear-time algorithm and
a randomized algorithm that runs in expected linear time.
Around the same time, Georgiadis et al.~\cite{GIK21} introduced another linear-time algorithm.
Since all these algorithms require the input graph to be 3-edge-connected,
 the input graph $G$ must be decomposed into its 3eccs in linear time first.
However, as a 3ecc is not a graph but a subset of vertices,
  The edges in $G$ that connect the vertices in the same 3ecc must be taken into consideration.
 The inclusion of these edges gives rise to the subgraph of $G$ induced by the 3eccs.
  Unfortunately, the induced subgraph may not be 3-edge-connected.
 This is because a path connecting two vertices in a 3ecc using edges outside the induced subgraph will vanish in the induced subgraph.
 As a result, while two vertices are connected by three edge-disjoint paths in $G$,
 they may  be connected by less than three edge-disjoint paths in the induced subgraph.
 Hence, new edges must be added to the induced subgraphs to take those vanished paths into consideration.
These new edges are identified by the following theorem based on~\cite{D92}.

\begin{thm}\label{auxiiary-2}
 Let $G=(V,E)$ be a connected graph.
 For each 3ecc of $G$, replace every 2-cut $\{(v,w),(\ddot{v},\ddot{w})\}$ in $G$, where
$v$ and $\ddot{v}$ are vertices in the 3ecc such that $v \neq \ddot{v}$ with a new edge $ (v,\ddot{v})$.
Let the resulting graph be $\acute{G}$.
 Then,
$\forall x,y \in V, x \overset{3}{\equiv}_{G} y$ if and only if  $x \overset{3}{\equiv}_{\acute{G}} y$.
\end{thm}

 \begin{figure}
 \centering
 \includegraphics[width=4.2in]{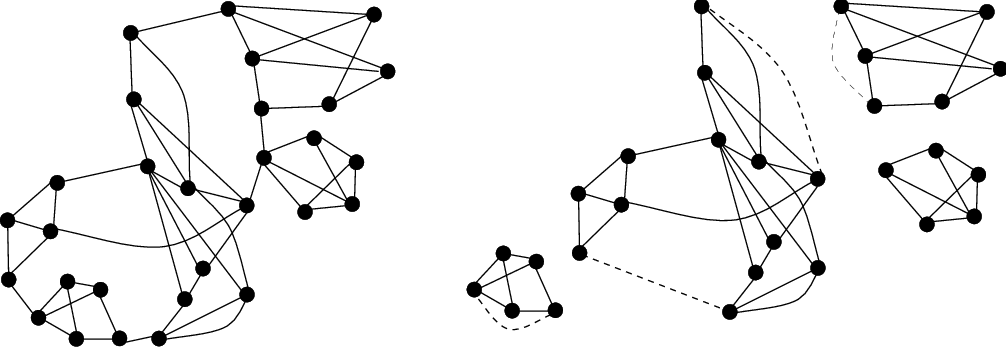}
 \caption[ ]{A graph and its auxiliary 3ecc subgraphs; dotted lines are auxiliary edges }
 \end{figure}\label{Fig1.eps}

The edges $(v,\ddot{v})$ and $(w,\ddot{w})$ are called the \emph{auxiliary edges}
corresponding to the 2-cut $\{(v,w),(\ddot{v},\ddot{w})\}$.
The subgraph induced by a 3ecc and the auxiliary edges added to it form an
\emph{auxiliary 3ecc subgraphs} of $G$~\cite{GIK21}.
To determine the 4-edge-connected components,
both Nadara et al. and Georgiadis et al. presented a linear-time algorithm for
constructing the auxiliary 3ecc subgraphs based on~\cite{D92}. We outline them as follows:

\singlespacing

\noindent\hrulefill

\noindent  \textbf{Algorithm} \texttt{Auxiliary-3ecc-subgraphs}
\footnotesize
of Georgiadis et al.~\cite{GIK21}\normalsize

\vspace{-6pt }
\noindent\hrulefill

\footnotesize

\noindent \textbf{Input:} An undirected multigraph $G=(V,E)$

\noindent \textbf{Output:} The auxiliary subgraphs of the 3ecc of $G$

\noindent
\textbf{begin}

\noindent
1. Compute the connected components of $G$ using breadth-first search;

\noindent
2. \textbf{for each} connected component \textbf{do}

   compute the 1-cuts using the algorithm of Tarjan~\cite{Tar72};

   compute  the 2eccs based on the 1-cuts;

\noindent
3. \textbf{for each}  2ecc  $H$ of $G$ \textbf{do}

   compute the 2-cuts in $H$ using the algorithm of Georgiadis et al.~\cite{GGIK21};

    compute  the 3eccs based on the 2-cuts;

\noindent
4. \textbf{for each}   2ecc $H$ of $G$ \textbf{do}

let the 3eccs be $C_1, C_2, \ldots, C_{\omega}$.

Construct the  auxiliary graphs for $C_i, 1 \le i \le \omega,$ as follows:

$(i)$  Shrink the induced subgraph of $C_i$ into a  vertex $\tilde{C}_i$,
resulting in a graph $\tilde{G}(\tilde{V},\tilde{E})$, where
$\tilde{V} = \{\tilde{C}_i \mid 1 \le i \le \omega \}$, and

\hspace{10pt} every edge in $\tilde{E}$ corresponds to an edge of a 2-cut in $H$ and lies on a unique cycle.

$(ii)$  \textbf{for each}   $\tilde{C}_i, 1 \le i \le \omega$ \textbf{do}

\hspace{25pt}  \textbf{for every} two edges, $(\tilde{C}_i, \tilde{C}_j), (\tilde{C}_i,\tilde{C}_{\ell})$ lying on the same cycle \textbf{do}

\hspace{35pt} let
$(\tilde{C}_i, \tilde{C}_j)$ and $(\tilde{C}_i,\tilde{C}_{\ell})$ correspond to
cut-edge $(x,y)$ and $(x',y')$ in $H$, respectively, and $x,x' \in C_i$;

\hspace{35pt} \textbf{if} $(x \neq x')$ \textbf{then} add edge $(x,x')$ to the induced subgraph of $C_i$;

\noindent
\textbf{end}

\normalsize
\noindent\hrulefill


\noindent\hrulefill

\noindent  \textbf{Algorithm} \texttt{Auxiliary-3ecc-subgraphs}
\footnotesize
of Nadara et al.~\cite{NRSS21}\normalsize

\vspace{-6pt }
\noindent\hrulefill

\footnotesize

\noindent \textbf{Input:} An undirected multigraph $G=(V,E)$

\noindent \textbf{Output:} The auxiliary subgraphs of the 3ecc of $G$

\noindent
\textbf{begin}

\noindent
1. Compute the connected components of $G$ using breadth-first search;

\noindent
2. \textbf{for each} connected component \textbf{do}

   compute the 2-vertex-connected components using the algorithm of Tarjan~\cite{Tar72};

\noindent
3. \textbf{for each}  2-vertex-connected  component \textbf{do}

    compute  the 3eccs using the 3-vertex-connected component algorithm of Hopcroft et al.~\cite{HT73}
    and

\hspace{88pt}    the reduction algorithm of Galil et al.~\cite{GI91};

\noindent
4. Same as Step 4 of the above \textbf{Algorithm} \texttt{Auxiliary-3ecc-subgraphs}
\footnotesize
of Georgiadis et al. \normalsize

\noindent
\textbf{end}

\normalsize
\noindent\hrulefill

\doublespacing

Recently, Tsin~\cite{T23} presented a linear-time certifying  algorithm for 3-edge-connectivity.
The algorithm produces, in addition to other outputs,
a Mader construction sequence for each 3ecc.
  The edges of the Mader construction are precisely the edges of the auxiliary 3ecc subgraphs for the same 3ecc.
 Hence, the algorithm provides another way to construct the auxiliary subgraphs for the 3eccs as follows.

\singlespacing

\noindent\hrulefill

\noindent  \textbf{Algorithm} \texttt{Auxiliary-3ecc-subgraphs}
\footnotesize
based on Tsin~\cite{T23}\normalsize

\vspace{-6pt }
\noindent\hrulefill

\footnotesize

\noindent \textbf{Input:} An undirected multigraph $G=(V,E)$

\noindent \textbf{Output:} The auxiliary graphs of the 3ecc of $G$

\noindent
\textbf{begin}

\noindent
1. Compute the connected components of $G$ using depth-first search;

\noindent
2.   \textbf{for each} connected component \textbf{do}

construct a Mader construction sequence for each 3ecc using the algorithm of Tsin~\cite{T23};

\noindent
\textbf{end}

\normalsize
\noindent\hrulefill

\doublespacing

Since the algorithm of Tsin~\cite{T23} performs only one depth-first search over each connected component of the input graph and
does not compute the 1-cuts, 2-cuts, 2eccs, 3eccs, 2-vertex-connected components, or  3-vertex-connected components,
 \textbf{Algorithm} \texttt{Auxiliary-3ecc-subgraphs}
\footnotesize
based on Tsin~\cite{T23}\normalsize
~is simpler than the two preceding algorithms.

To construct a Mader construction sequence, Tsin's algorithm~\cite{T23} partitions the edges of each connected component  into a collection of paths, and
 adds the paths to the initially empty sequence in a particular order.
 To use the algorithm to construct the auxiliary 3ecc subgraphs,
  since we are only interested in the edge set of the Mader construction sequence,
  not the order in which the edges are added to the sequence.
   we shall simplify the algorithm to produce a much simpler linear-time algorithm for constructing the
  auxiliary subgraphs of the 3eccs.

\vspace{-6pt}

\section{Basic definitions and facts}

\vspace{-6pt}

An undirected graph is represented by $G=(V,E)$, where $V$ is the vertex set and $E$ is the edge set.
An edge $e$ with $u$ and $v$ as end-vertices is represented by $e=(u,v)$.
The graph is a \emph{multigraph} if it contains \emph{parallel edges} (two or more edges sharing the same pair of end vertices).
The \emph{degree} of a vertex $u$ in $G$, denoted by $deg_G(u)$, is the number of edges with $u$ as an end-vertex.
A \emph{path} $P$ in $G$ is a sequence of alternating vertices
   and edges,
   $u_0 e_1 u_1 e_2 u_2 \ldots e_k u_k$, such that $u_i \in V, 0 \le i
   \le k,$ $e_i  = (u_{i-1}, u_i), 1 \le i \le k$, where $u_i, 0 \le i
   \le k,$ are distinct with the exception that $u_0$ and $u_k$ may be identical.
    The edges $e_i, 1 \le i \le k,$ could be omitted if no
    confusion could occur as a result.
     The path is a \emph{null path} if $k=0$ and is a \emph{cycle} if  $u_0 = u_k$.
     The path is called an $u_0-u_k$ \emph{path}.
    If the path $P$ is given an orientation from $u_0$ to $u_k$, then $u_0$ is the \emph{source}, denoted by $s(P)$, and $u_k$ is the \emph{sink}, denoted by $t(P)$, of $P$ and the path $P$ is also represented by $u_0 \rightsquigarrow_G u_k$.
         The graph $G$ is \emph{connected} if $\forall u,v \in V$, there is a $u-v$ path
     in it;  it is \emph{disconnected} otherwise.
 A graph $G' =(V', E')$ is a \emph{subgraph} of $G$ if $V' \subseteq V$ and $E' \subseteq E$.
  Let $U \subseteq V$, the \emph{subgraph of $G$ induced by $U$}, denoted by $G_{\langle U \rangle}$, is the maximal subgraph of $G$ whose vertex set is $U$.
     Let $D \subseteq E$, $G \setminus D$ denotes the graph resulting from $G$ after
     the edges in $D$ are removed.
%

  \emph{Depth-first search} (abbreviated \emph{dfs}) augmented with vertex labeling is a powerful graph
  traversal technique~\cite{Tar72}.
     When a $\mathit{dfs}$ is performed over a graph, each
     vertex $w$
     is assigned a \emph{depth-first number}, $dfs(w)$, such that
     $dfs(w)=k$ if vertex $w$ is the $k$th vertex visited by the search for the first time.
      The search also partitions the edge set into two types of edges,
      \emph{tree-edge} and \emph{back-edge}
and gives each edge an orientation.
      With the orientation taken into consideration, a tree-edge $e = (u,v)$ is denoted by $u \rightarrow v$ where  $dfs(u) < dfs(v)$
      and a back-edge $e = (u,v)$ is denoted by $u \curvearrowright v$ or $v \curvearrowleft u$, where $dfs(v) < dfs(u)$.
      In the former case,  $u$ is the \emph{parent} of
       $v$, denoted by $u=parent(v)$, while $v$ is a \emph{child} of $u$.
      In the latter case, $u \curvearrowright v$ is an \emph{incoming back-edge} of $v$ and
       an \emph{outgoing back-edge} of $u$.
In either case, $u$ is the \emph{tail}, denoted by $s(e)$, while $v$ is the \emph{head}, denoted by $t(e)$, of the edge.
         The tree edges form a directed spanning tree $T = (V, E_T)$ of
$G$ rooted at the vertex $r$ where the search begins.
          A path from vertex $u$ to vertex $v$ in $T$ is denoted
          by $u \rightsquigarrow_T v$.
 It is also called an $u-v$ \emph{tree-path}.
   Vertex $u$ is an \emph{ancestor} of vertex $v$, denoted by $u \preceq v$, if and only if $u$ is a vertex on $r \rightsquigarrow_T v$.
   Vertex $u$ is a \emph{proper ancestor} of $v$, denoted by $u \prec v$, if $u \preceq v$ and $u \neq v$.
 Vertex $v$ is a (\emph{proper}) \emph{descendant} of vertex $u$ if and only if vertex $u$ is an  (proper) ancestor of vertex $v$.
   When a $\mathit{dfs}$ reaches a vertex $u$, vertex $u$ is
   called the \emph{current vertex} of the search.
    The \emph{subtree} of $T$ \emph{rooted at vertex} $w$, denoted by
    $T_w$, is the subtree containing all the descendants of $w$.
The \emph{palm tree rooted at} $w$, $T_w^{\curvearrowright}$, is
$T_w$ including the back-edges who tails are in $T_w$~\cite{Tar72}.

\noindent $\forall w \in V, lowpt(w)  = \min (  \{dfs(w)\} \cup
\{ dfs(u) \ | \ (w \curvearrowright u) \in E \setminus E_T \}  \cup \{ lowpt(u) \ | \ (w
\rightarrow u) \in  E_T \}  )$.
Specifically, $lowpt(w)$ is the smallest $\mathit{dfs}$ number of a  vertex reachable from $w$ via a (possibly null)  tree-path followed by a back-edge~\cite{Tar72}.




\vspace{-6pt}

\section{A linear-time algorithm for generating the auxiliary 3ecc subgraphs}

\vspace{-6pt}

\hspace{15pt}
Since our algorithm is based on
 \textbf{Algorithm} 3-edge-connectivity of~\cite{T07},
 we first review the algorithm.

The key idea underlying the algorithm is to use the following graph contraction operation, called \emph{absorb-eject},
to gradually transform the input graph $G$ into an edgeless  graph
of which each vertex corresponds to a distinct 3ecc of $G$.

\noindent \textbf{Definition:}
  Let $G' = (V', E')$ and $e=(w,u) \in E'$ such that:

$(i)$ $deg_{G'}(u) = 2$, or
$(ii)$ $e$ is not a cut-edge (implying $w$ and $u$ are 3-edge-connected).

   Applying the \emph{absorb-eject} operation to $e$ at $w$ results in the graph $G'/e = (V'', E'')$, where

\hspace{20pt}   $V'' = \left\{
           \begin{array}{ll}
             V',                 & \hbox{if $deg_{G'}(u) = 2$;} \\
             V' \setminus \{u\}, & \hbox{if $e$ is not a cut-edge, }
           \end{array}
         \right.
  $

\vspace{6pt}
\hspace{20pt}  $E'' = E' \setminus E_u \cup E_{w^+}$,
where $E_u$ is the set of edges incident on $u$ in $G'$ and
      $E_{w^+} = \{ f' = (w,z) \mid \exists f = (u, z) \in E_u$, for some $z \in V' \backslash \{w\} \}$.

The edge $f'=(w,z)$ is called an embodiment of the edge $f = (u,z)$.
In general, an \emph{embodiment} of an edge $f$ is the edge $f$ itself, or
an edge created to replace $f$ as a result of applying the absorb-eject operation,
 or an embodiment of an embodiment of $f$.

 The effect of applying the absorb-eject operation is illustrated in Figure 2.
In Case $(i)$,
  $deg_{G'}(u)=2$.
  Vertex $w$ absorbs edge $e$ and ejects vertex $u$ making it an isolated vertex in $G'/e$.
  Edge $(u,d)$ is then replaced by its embodiment $(w,d)$.
In Case $(ii)$, $e$ is not a cut-edge.
Vertex $w$ absorbs edge $e$ and vertex $u$.
Moreover, all edges $f = (u,z)$ incident on $u$, except $e$, are replaced by their embodiment $f'=(w,z)$.

 \begin{figure}
 \centering
 \includegraphics[width=4.2in]{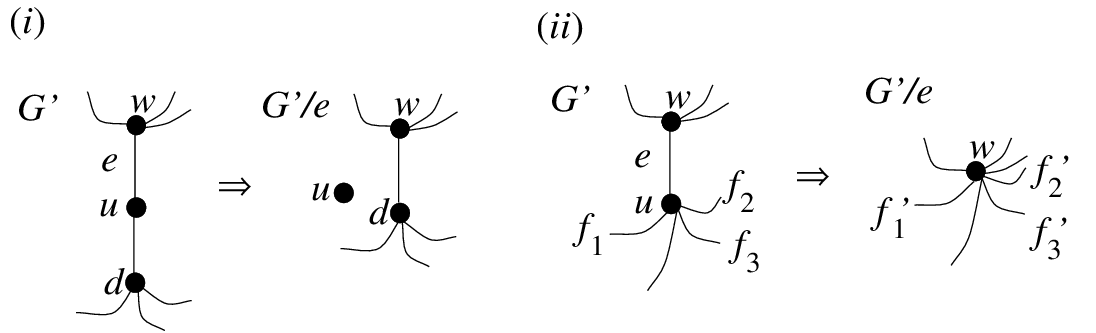}
 \caption[ ]{The absorb-eject operation: $(i)$ $deg_{G'}(u) = 2$;
 $(ii)$ edge $(w,u)$ is not a cut-edge. }
 \end{figure}\label{Fig2.eps}

Starting with the input graph $G = (V, E)$, using the absorb-eject operation, the graph is
gradually transformed so that vertices that have been confirmed to be belonging to the same $3ecc$ are merged into one vertex, called a \emph{supervertex}.
Each supervertex is represented by a vertex $w \in V$ and a set $\sigma(w) (\subseteq V)$
consisting of vertices that have been confirmed
to be belonging to the same $3ecc$
as $w$. Initially, each vertex $w$ is regarded as a supervertex with $\sigma(w) = \{w\}$.
When two adjacent supervertices $w$ and $u$ are known to be belonging to the same
$3ecc$, the absorb-eject operation is applied to have one of them, say $w$, absorbing the other resulting in $\sigma(w) := \sigma(w) \cup \sigma(u)$.
When a supervertex containing all  vertices of a $3ecc$ is formed,
it must have \emph{degree one or two} in the transformed graph (corresponding to a 1-cut or a 2-cut is found).
 Then, the absorb-eject operation is applied to an adjacent supervertex
 to separate (\emph{eject}) the supervertex from the graph making it an isolated vertex.
At the end, the graph is transformed into a collection
of isolated supervertices each of which contains the vertices of a distinct $3ecc$ of $G$.

During the $\mathit{dfs}$, at each vertex $w \in V$,
   when the adjacency list of $w$ is completely processed,
 let $u$ be a first child of $w$ encountered such that $lowpt(w) = lowpt(u)$.
   The palm tree rooted at $u$, $T^{\curvearrowright}_u$, with $w \rightarrow u$ have been transformed into a path
of supervertices, $\mathcal{P}_w: (w=)w_0 w_1 w_2 \ldots w_k$, called the $w$-\emph{path}, where
$w_1 = u$ if and only if $u$ was not ejected, and
a set of isolated supervertices each of which corresponds to a $3ecc$ residing in $T^{\curvearrowright}_u$.
   For each of the remaining child vertex $u'$ of $w$,
   its palm tree $T^{\curvearrowright}_{u'}$  has been transformed into a set of isolated supervertices,
each corresponding to a $3ecc$ residing in $T^{\curvearrowright}_{u'}$, and
   all the vertices of $T^{\curvearrowright}_{u'}$ not belonging to any of the 3eccs have been absorbed by $w$.
   The $w$-path has the following properties:

\vspace{-3pt}
\begin{lem} \label{Tsin07}

\emph{[Lemma 6 of~\cite{T07}]} ~
    Let $\mathcal{P}_w: (w=)w_0 w_1 w_2 \ldots w_k$ and
    $v = parent(w)$.                                          

 \begin{description}
    \vspace{-6pt}
   \item[$(i)$] $deg_{\hat{G}_w}(w_0) \ge 1$  and $deg_{\hat{G}_w}(w_i) \ge 3, 1 \le i \le k$, where $\hat{G}_w$ is the graph to which $G$ has been transformed;

   \vspace{-6pt}
   \item[$(ii)$]  for each back-edge $f = (w_i \curvearrowright x), 0 \le i \le k,$
      $x \preceq v $ (i.e. $x$ lies on the $r \rightsquigarrow_T v$ tree-path);

    \vspace{-6pt}
   \item[$(iii)$] $\exists (w_k \curvearrowright z)$ such that $dfs(z) = lowpt(w)$. 
 \end{description}
\end{lem}

\vspace{-6pt}
 Specifically, on the $w$-path, the degree of every supervertex is at least \emph{three} (else, the supervertex would have been ejected) except that of $w_0$;
 there is no back-edge connecting two supervertices on the $w$-path
 (else, the supervertices lying in between the head and tail of the back-edge would have been absorbed by the head);
 the  supervertex $w_k$ has an outgoing back-edge reaching the vertex whose \emph{dfs} number is $lowpt(w)$.

 The $w$-path is constructed as follows.
 Initially, the $w$-path is the null path $w$ which is the \emph{current} $w$-\emph{path},
 and $lowpt(w) = dfs(w)$.
 When the \emph{dfs} backtracks from a child $u$, let
 $\mathcal{P}_u: (u=)u_0 u_1 u_2 \ldots u_h$ be the $u$-\emph{path},
 and
 $\hat{G}_u$ be the graph to which $G$ has been transformed at that point of time.

 If $deg_{\hat{G}_u}(u) = 1$, then $(w,u)$ is a 1-cut (or \emph{bridge}) and $\sigma(u)$ is a $3ecc$ of $G$.
 Edge  $(w,u)$ is removed making $u$  an isolated supervertex.
 The current $w$-path remains unchanged.
 If $deg_{\hat{G}_u}(u) = 2$, then $\{(w \rightarrow u), (u \rightarrow u_1)\}$ or
  $\{(w \rightarrow u), (u \curvearrowright z)\}$,
  where $dfs(z) = lowpt(u)$, is a 2-cut
 implying $\sigma(u)$ is a $3ecc$ of $G$.
   The absorb-eject operation is applied  to eject  $u$ from the $u$-path making it an isolated supervertex and the
   $u$-path is shorten to $u_1 u_2 \ldots u_k$ in the former case  or vanishes in the latter case.
   Now, if $lowpt(w) \le lowpt(u)$,
      the vertices in the supervertices on the $u$-path must all belong to the same $3ecc$ as $w$.
 The supervertices are thus absorbed by $w$ through a sequence of absorb-eject operations.
 If $lowpt(w) > lowpt(u)$,  the vertices in the supervertices on the current $w$-path must all belong to the same $3ecc$ as $w$;
 the supervertices are thus absorbed by $w$.
 Moreover, $lowpt(w) := lowpt(u)$,
and if the $u$-path vanishes,
the current $w$-path becomes $\mathcal{P}_w: w$.
Otherwise, the $u$-path after extended to include $w$ becomes the current $w$-path.

   On encountering an outgoing back-edge $(w \curvearrowright z)$, with $dfs(z) < lowpt(w)$,
vertex $w$ absorbs the current $w$-path as the supervertices on it belong to the same $3ecc$ as $w$;
$lowpt(w)$ and the $w$-path are  updated to $dfs(z)$ and $\mathcal{P}_w: w$, respectively.

On encountering an incoming back-edge $(w \curvearrowleft x)$,
if $\exists h, 1 \le h \le k,$ such that $x \in \sigma(w_{h})$ on the current $w$-path,
the vertices in $\sigma(w_i), 1 \le i \le h$,
  must all belong to the same $3ecc$ as $w$.
   The supervertices $w_i, 1 \le i \le {h},$ are thus absorbed by $w$
    and
the current $w$-path is shortened to $w w_{h+1} w_{h+2} \ldots w_k$.

When the adjacency list of $w$ is completely processed,
if $w \neq r$, the current $w$-path becomes the $w$-path,
and
the $\mathit{dfs}$ backtracks to the parent vertex of $w$.
Otherwise, the input graph $G$ has been transformed into a collection of isolated supervertices each of which contains the vertices of a
distinct $3ecc$ of $G$.
A complete example is given in~\cite{T07}, pp.132-133.

The term $lowpt(w), w \in V,$ plays an important role in~\cite{T07}  for finding  the 3eccs.
It is the $\mathit{dfs}$ number of the  vertex closest to the root $r$ that can be reached from $w$ by a possibly null tree-path followed by an outgoing back-edge.
Since our algorithm is an extension of that in~\cite{T07},
$lowpt(w)$ continues to play an important role.
However, while a 3ecc  is edgeless, an auxiliary subgraph is not.
Therefore, we must take the path that reaches the vertex with $\mathit{dfs}$ number
$lowpt(w)$ into consideration.
Such a path can be determined based on an ordering of the back-edges as follows.

    Let $T=(V,E_{T})$  be a $\mathit{dfs}$ tree of  $G$.
Using the $\mathit{dfs}$ numbers,
the back-edges can be ranked as follows.

\noindent   \textbf{Definition:}
 Let $(p \curvearrowright q)$ and $(x \curvearrowright y)$ be two back-edges.
     Then   $(p \curvearrowright q)$  is  \emph{lexicographically smaller than} $(x \curvearrowright y)$, denoted by $(p \curvearrowright q) \lessdot (x \curvearrowright y)$,
  if and only if

  (i) $dfs(q)  < dfs(y)$, or

  (ii) $q = y$
        and
          $dfs(p)  < dfs(x)$
        such that  $p$  is not an ancestor of  $x$, or

  (iii) $q = y$  and  $p$  is a proper descendant of $x$.   

Parallel back-edges are ranked in arbitrary order among themselves.


  Since every tree-edge is the parent edge of a unique vertex,
   every tree-edge can be represented by $(parent(w) \rightarrow w)$.
    Using the back-edges, we can partition the edges of $G$, excluding the 1-cuts,
  into edge-disjoint paths such that
every path contains exactly one back-edge as follows:
   for each tree-edge   $(parent(w) \rightarrow w)$ that is not a 1-cut,
 we associate with it the back-edge $(x \curvearrowright y)$
  with the  lowest rank in lexicographical order such that
      $w \preceq x$     
  while $y \prec w$.     


  It is easily verified that the back-edge $(x \curvearrowright y)$ and all the
tree-edges associated with it form a path
 $yxv_{1} \cdots v_{p}w$   such that
$w \rightsquigarrow_T x$. 
  The path is called an \emph{ear}, denoted by $P_{x \curvearrowright y}: yxv_{1} \cdots v_{p}w$ or
$w \rightsquigarrow_T x \curvearrowright y$.
  Since $P_{x \curvearrowright y}$ consists of a possibly null tree-path $w \rightsquigarrow_T x$ followed by an outgoing back-edge $(x \curvearrowright y)$, and
   $y$ is the vertex closest to the root $r$ that can be reached from $w$ through
    such a path,
     $lowpt(w) =  dfs(y)$.

     The back-edge $(x \curvearrowright y)$ associated with the parent edge of $w$, denoted by
     $ear(w)$, is computed during the $\mathit{dfs}$ based on the following recursive definition:

$ear(w) =
    \min_{\lessdot} (\{ f \mid f =(w \curvearrowright u) \in E \setminus E_T \} \cup
     \{ ear(u) \mid f = (w \rightarrow u) \in E_T \}  \cup
     \{ \overset{\curvearrowright}{\infty} \} )$,
where
$\overset{\curvearrowright}{\infty}$ is the initial value of $ear(w)$ such that
$ (x \curvearrowright y) \lessdot \overset{\curvearrowright}{\infty}$ if
   $(y \prec w) \wedge (w \preceq x)$ and
$\overset{\curvearrowright}{\infty}  \lessdot  (x \curvearrowright y)$ if
   $w \preceq y,  \forall (x \curvearrowright y) \in E \backslash E_T$.
It is easily verified that
$ear(w) = \overset{\curvearrowright}{\infty}$
 iff $w = r$ or the parent edge of $w$ is a 1-cut.
Since the graph is transformed during the $\mathit{dfs}$,
at each vertex $w$, whenever the $\mathit{dfs}$ backtracks from a child $u$,
the palm tree rooted at $u$, $T_u^{\curvearrowright}$, might have been transformed.
As a result, $ear(w)$ is computed based on the transformed palm tree.

\noindent
\textbf{Definition:}
The \emph{transformed palm tree rooted at} $w$,
 denoted by $\dot{T}_w^{\curvearrowright}$,
is $T_w^{\curvearrowright}$ with
all the 1-cuts removed and
all the 2-cuts replaced by their corresponding auxiliary edges,
disregarding the auxiliary 3ecc subgraphs.
Let $\dot{T}_u^{\curvearrowright} \cup (w \rightarrow u)$ denote
$\dot{T}_u^{\curvearrowright}$ extended to include $(w \rightarrow u)$.
The \emph{extended transformed palm tree of} $u$,
 denoted by $\ddot{T}_u^{\curvearrowright}$,
is $\dot{T}_u^{\curvearrowright} \cup (w \rightarrow u)$ with
all the 1-cuts removed and
all the 2-cuts replaced by their corresponding auxiliary edges,
disregarding the auxiliary 3ecc subgraphs.
(Figure 3)

It is easily verified that
$\dot{T}_w^{\curvearrowright} =
\{ f \mid f =(w \curvearrowright u) \in E \setminus E_T \} \cup
\bigcup \{ \ddot{T}_u^{\curvearrowright} \mid (w \rightarrow u) \in E_T \}$.
Let $ear_{\ddot{T}_u^{\curvearrowright}}(w)$
denote $ear(w)$ in $\ddot{T}_u^{\curvearrowright}$.
Then,

 $ear(w) =
ear_{\dot{T}_w^{\curvearrowright}}(w)
= \min_{\lessdot} (\{ f \mid f =(w \curvearrowright u) \in E \setminus E_T \} \cup
     \{ ear_{\ddot{T}_u^{\curvearrowright}}(w) \mid (w \rightarrow u) \in E_{T} \}  \cup
     \{ \overset{\curvearrowright}{\infty} \} )$.

\vspace{-6pt}
 \begin{figure}[hbt!]
 \centering
 \includegraphics[width=6in]{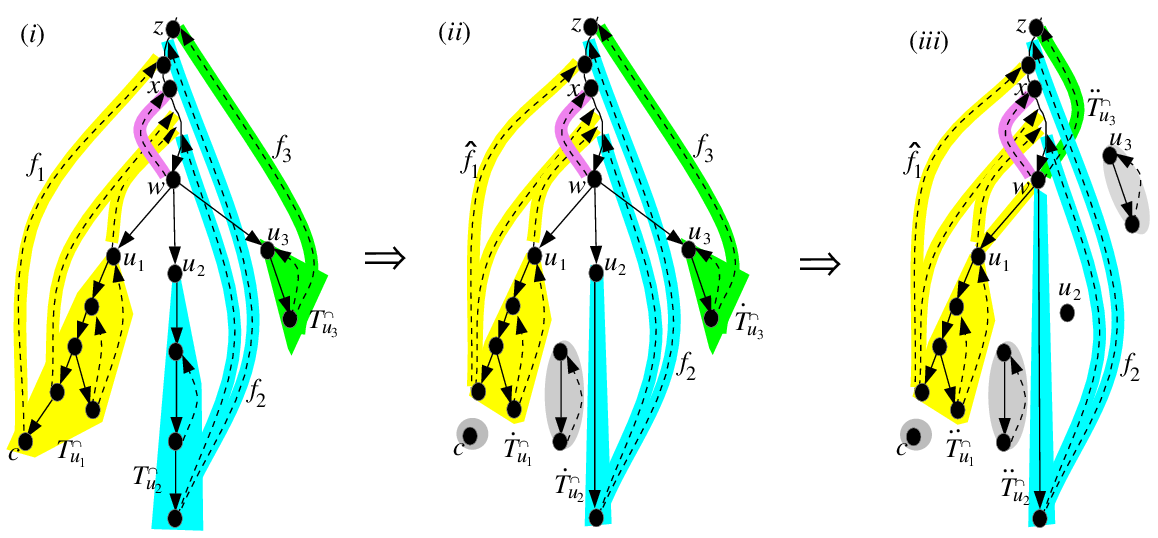}
 \caption[ ]
 { $(i)$ Palm trees: $T_{u_1}^{\curvearrowright}, T_{u_2}^{\curvearrowright}, T_{u_3}^{\curvearrowright}$;
 $(ii)$  Transformed palm trees:
 $\dot{T}_{u_1}^{\curvearrowright}, \dot{T}_{u_2}^{\curvearrowright}, \dot{T}_{u_3}^{\curvearrowright}$;
 $(iii)$  Extended transformed palm trees:
   $\ddot{T}_{u_1}^{\curvearrowright}, \ddot{T}_{u_2}^{\curvearrowright}, \ddot{T}_{u_3}^{\curvearrowright}$, and transformed palm tree $\dot{T}_w^{\curvearrowright}$.

 $ear(w) =
ear_{\dot{T}_{w}^{\curvearrowright}}(w)
 = \min_{\lessdot}\{ ear_{\ddot{T}_{\tilde{x}}^{\curvearrowright}}(\tilde{x}), ear_{\ddot{T}_{u_1}^{\curvearrowright}}(w),
 ear_{\ddot{T}_{u_2}^{\curvearrowright}}(w),
 ear_{\ddot{T}_{u_3}^{\curvearrowright}}(w)$
$
= \min_{\lessdot}\{ (w \curvearrowright x),ear_{\ddot{T}_{u_1}^{\curvearrowright}}(w),$

$
 ear_{\ddot{T}_{u_2}^{\curvearrowright}}(w),
(w \curvearrowright t(ear(u_3))) \}$
$= \min_{\lessdot}\{ (w \curvearrowright x), \hat{f_1}, f_2, (w \curvearrowright z) \}
= (w \curvearrowright z).
$

}
 \end{figure}\label{Fig3}

\vspace{-9pt}
Next, we shall determine the auxiliary edges correspond to each 2-cut.
%


\vspace{-6pt}
\begin{lem}\label{2cut-iff-2}

When the $\mathit{dfs}$ backtracks from vertex
$u$ to  vertex $w$,
Let $\mathcal{P}_u: (u=) u_0 u_1 u_2 \ldots u_h (h \ge 0)$ be the $u$-path.
If $deg_{\hat{G}_u}(u) = 2$, and

\vspace{-9pt}
\begin{description}
  \item[$(i)$] $\{(w \rightarrow u), (u \rightarrow u_1)\}$ is a 2-cut.
   The corresponding 2-cut in $\dot{T}_w^{\curvearrowright}$ is
   $\{(w,u), (parent_{\dot{T}_u^{\curvearrowright}}(u_1) \rightarrow u_1)\}$,
whose corresponding auxiliary edges are
   $\{ (w, u_1), (u, parent_{\dot{T}_u^{\curvearrowright}}(u_1) \}$,
   where $parent_{\dot{T}_u^{\curvearrowright}}(u_1)$ is the parent vertex of $u_1$ in $\dot{T}_u^{\curvearrowright};$

\vspace{-6pt}
  \item[$(ii)$]  $\{(w \rightarrow u), (u \curvearrowright z)\}$ is a 2-cut (i.e., $h = 0$).
    The corresponding 2-cut in $\dot{T}_w^{\curvearrowright}$  is
    $\{(w,u), ear(u)\}$,
whose corresponding auxiliary edges are
   $\{ (w \curvearrowright t(ear(u)) ),
 (u, s(ear(u)) )\}$, where $z = t(ear(u))$.

\end{description}

\end{lem}

\vspace{-9pt}
\noindent \textbf{Proof:}
\vspace{-9pt}

\begin{description}
  \item[$(i)$]
  The $u$-path contains $u$ and $u_1$ implies that there is an $u$-$u_1$ path in $\dot{T}_u^{\curvearrowright}$.
  Since $u_1$ remains on the $u$-path when the $\mathit{dfs}$ backtracks from $u$ to $w$,
  vertex $u$ must have absorbed the path
   $u \rightsquigarrow_{\dot{T}_u^{\curvearrowright}} parent_{\dot{T}_u^{\curvearrowright}}(u_1)$
   which implies that $(u \rightarrow u_1)$ is an embodiment of
   $(parent_{\dot{T}_u^{\curvearrowright}}(u_1) \rightarrow u_1)$.
   Hence,
   $\{(w,u), (parent_{\dot{T}_u^{\curvearrowright}}(u_1) \rightarrow u_1)\}$ is the 2-cut corresponding to $\{(w \rightarrow u), (u \rightarrow u_1)\}$
in $\dot{T}_w^{\curvearrowright}$
and
$\{(w, u_1), (u, parent_{\dot{T}_u^{\curvearrowright}}(u_1)\}$ are the corresponding auxiliary edges.

\vspace{-6pt}
  \item[$(ii)$]
  Since $u = u_0$,
   vertex $u$ must have absorbed  $V_{\dot{T}_{u}^{\curvearrowright}}$
   which includes $s(ear(u))$.
  Since $\{(w \rightarrow u), (u \curvearrowright z)\}$ is a 2-cut implies $deg_{\hat{G}_u}(u) = 2$,
  $ear(u)$ is the only back-edge
  with $u \preceq s(ear(u))$ and
  $t(ear(u)) \prec u$
   which implies that
  $(u \curvearrowright z)$ is an embodiment of $ear(u)$.
Therefore,
$z = t(ear(u))$.
  Moreover,
  $\{(w,u), ear(u)\}$ is the 2-cut
  corresponding to $\{(w \rightarrow u), (u \curvearrowright z)\}$ in $\dot{T}_w^{\curvearrowright}$
and
$\{ ( w \curvearrowright t(ear(u)) )$,
  $(u, s(ear(u))) \}$ are the corresponding auxiliary edges.
   \ \ \ \ \   $\blacksquare$

\end{description}

\begin{lem} \label{palm-tree}
Let $u$ be a child of $w$.
When the $\mathit{dfs}$ backtracks from $u$ to $w$,
let
$P_u: u(=u_0) u_1  \ldots u_h (h \ge 0)$ and
$(w, u)$ is not a 1-cut. Then,

\vspace{3pt}
$\ddot{T}_u^{\curvearrowright} =
\left\{
  \begin{array}{ll}
    (w \curvearrowright t(ear(u))), & \hbox{if $(w, u)$ is a cut-edge in $\dot{T}_u^{\curvearrowright} \cup (w \rightarrow u)$ and $h = 0$;} \\
    \dot{T}_{u_1}^{\curvearrowright} \cup (w \rightarrow u_1), & \hbox{if $(w, u)$ is a cut-edge in $\dot{T}_u^{\curvearrowright} \cup (w \rightarrow u)$ and $h > 0$;} \\
    \dot{T}_u^{\curvearrowright} \cup (w \rightarrow u), & \hbox{otherwise.}
  \end{array}
\right.
$.
\end{lem}

\noindent
\textbf{Proof:}

$(a)$
If $(w,u)$ is a cut edge in $\dot{T}_u^{\curvearrowright} \cup (w \rightarrow u)$ and $h = 0$,
the 2-cut in $\hat{G}_u$ is $\{ (w \rightarrow u), (u \curvearrowright z) \}$.
By Lemma~\ref{2cut-iff-2}$(ii)$, the corresponding 2-cut in $\dot{T}_w^{\curvearrowright}$, hence in $\dot{T}_u^{\curvearrowright} \cup (w \rightarrow u)$, is $\{ (w \rightarrow u), ear(u)\}$ whose corresponding auxiliary edges are
   $\{ (w \curvearrowright t(ear(u)) ), (u, s(ear(u)) )\}$,
where $z= t(ear(u))$.
It follows that $\dot{T}_u^{\curvearrowright} \cup (w \rightarrow u)$ is transformed to
$(w \curvearrowright t(ear(u)))$
after the 2-cut and the auxiliary subgraph of
$\sigma(u)$ are removed.
Hence,
$\ddot{T}_u^{\curvearrowright} =
(w \curvearrowright t(ear(u)))$.

$(b)$
If $(w,u)$ is a cut edge in $\dot{T}_u^{\curvearrowright} \cup (w \rightarrow u)$ and $h > 0$,
the 2-cut in $\hat{G}_u$ is $\{ (w \rightarrow u), (u \rightarrow u_1) \}$.
By Lemma~\ref{2cut-iff-2}$(i)$, the corresponding 2-cut in $\dot{T}_w^{\curvearrowright}$, hence in $\dot{T}_u^{\curvearrowright} \cup (w \rightarrow u)$, is
$\{(w,u), (parent_{\dot{T}_u^{\curvearrowright}}(u_1) \rightarrow u_1)\}$,
   whose corresponding auxiliary edges are $\{ (w, u_1), (u, parent_{\dot{T}_u^{\curvearrowright}}(u_1) \}$.
After the 2-cut and the auxiliary subgraph of
$\sigma(u)$ are removed,
$\dot{T}_u^{\curvearrowright} \cup (w \rightarrow u)$ is transformed to
$\dot{T}_{u_1}^{\curvearrowright} \cup (w \rightarrow u_1)$.
Hence,
$\ddot{T}_u^{\curvearrowright} =
\dot{T}_{u_1}^{\curvearrowright} \cup (w \rightarrow u_1)$.

$(c)$
If $(w,u)$ is not a cut edge in $\dot{T}_u^{\curvearrowright} \cup (w \rightarrow u)$,
then as no absorb-eject operation is applied to $(w \rightarrow u)$,
$\dot{T}_u^{\curvearrowright} \cup (w \rightarrow u)$ is intact.
Hence,
$\ddot{T}_u^{\curvearrowright} =
\dot{T}_u^{\curvearrowright} \cup (w \rightarrow u)$.
  \ \ \ \ \  $\blacksquare$

\vspace{-3pt}
\noindent
\begin{lem} \label{palm-tree-ear}
Let $u$ be a child of $w$ such that
$(w, u)$ is not a 1-cut. Then,

\vspace{9pt}
$ear_{\ddot{T}_u^{\curvearrowright}}(w)
= \left\{
    \begin{array}{ll}
      (w \curvearrowright t(ear(u))),
                & \hbox{if $\{(w,u), ear(u)\}$ \text{ is a 2-cut};} \\
      ear(u),   & \hbox{otherwise.}
    \end{array}
  \right.$

\end{lem}

\vspace{9pt}
\noindent
\textbf{Proof:}
Let $P_u: u (= u_0) u_1 \ldots u_h (h \ge 0)$ and
 $ear(u) = (x \curvearrowright z)$.

$(a)$
If $\{(w,u), ear(u)\}$ is a 2-cut,
then
$(w,u)$ is a cut edge in $\dot{T}_u^{\curvearrowright} \cup (w \rightarrow u)$ and $h = 0$.
By Lemma~\ref{palm-tree},
$\ddot{T}_u^{\curvearrowright} =
    (w \curvearrowright t(ear(u)))$
which implies that
$ear_{\ddot{T}_u^{\curvearrowright}}(w)
=   (w \curvearrowright t(ear(u))).$

$(b)$
If $\{(w,u), ear(u)\}$ is not a 2-cut
but $(w,u)$ is a cut-edge in $\dot{T}_u^{\curvearrowright} \cup (w \rightarrow u)$,
then
 $h > 0$.
By Lemma~\ref{palm-tree},
 $\ddot{T}_u^{\curvearrowright} =
\dot{T}_{u_1}^{\curvearrowright} \cup (w \rightarrow u_1)$.
Then,
 $ear(u) = (x \curvearrowright z)$ implies
 $u \rightsquigarrow_{\dot{T}_u^{\curvearrowright}}  \hspace{-3pt} u_1   \hspace{-3pt} \rightsquigarrow_{\dot{T}_u^{\curvearrowright}} \hspace{-3pt} x \curvearrowright z$
is the lexicographically smallest ear in $\dot{T}_u^{\curvearrowright}$ with $z$ closest to the root $r$
which implies that
 $u_1   \hspace{-3pt} \rightsquigarrow_{\dot{T}_{u_1}^{\curvearrowright}} \hspace{-3pt} x \curvearrowright z$
is the lexicographically smallest ear in $\dot{T}_{u_1}^{\curvearrowright}$ with $z$ closest to the root $r$
which in turn implies that
 $w \rightarrow u_1   \hspace{-3pt} \rightsquigarrow_{\dot{T}_{u_1}^{\curvearrowright}} \hspace{-3pt} x \curvearrowright z$
is the lexicographically smallest ear in $\dot{T}_{u_1}^{\curvearrowright} \cup (w \rightarrow u_1)$ with $z$ closest to the root $r$.
Hence,
$ear_{\ddot{T}_u^{\curvearrowright}}(w) = (x \curvearrowright z) \Rightarrow
ear_{\ddot{T}_u^{\curvearrowright}}(w) = ear(u)$.

$(c)$
If $(w,u)$ is not a cut-edge in $\dot{T}_u^{\curvearrowright} \cup (w \rightarrow u)$,
By Lemma~\ref{palm-tree},
 $\ddot{T}_u^{\curvearrowright} =
\dot{T}_{u}^{\curvearrowright} \cup (w \rightarrow u)$.
Then  $ear(u) = (x \curvearrowright z)$
implies $u \rightsquigarrow_{\dot{T}_u^{\curvearrowright}} \hspace{-3pt} x \curvearrowright z$
is the lexicographically smallest ear in $\dot{T}_u^{\curvearrowright}$ with $z$ closest to the root $r$
which implies that
$w \rightarrow u \rightsquigarrow_{\dot{T}_u^{\curvearrowright}} \hspace{-3pt} x \curvearrowright z$
is  the lexicographically smallest ear  in $\dot{T}_u^{\curvearrowright}\cup (w \rightarrow u)$ with $z$ closest to the root $r$.
Hence,
$ear_{\ddot{T}_u^{\curvearrowright}}(w) = (x \curvearrowright z) \Rightarrow
ear_{\ddot{T}_u^{\curvearrowright}}(w) = ear(u)$.
  \ \ \ \ \  $\blacksquare$

To simplify the notation, we  eliminate the back-edges by applying the following transformation to  $T_w^{\curvearrowright}$:
For each back-edge $(w \curvearrowright u)  \in E \setminus E_T$,
replace $(w \curvearrowright u)$ with a path consisting of two edges $(w \rightarrow \tilde{u})$ and $(\tilde{u} \curvearrowright u)$, where $\tilde{u}$ is a new fictitious vertex.
     It is easily verified that
     $ear_{\ddot{T}_{\tilde{u}}^{\curvearrowright}}(w) = (w \curvearrowright u)$.
Hence, $ear(w)$ can be defined as:

 $ear(w) =
\min_{\lessdot} (\{  ear_{\ddot{T}_{\tilde{u}}^{\curvearrowright}}(w) \mid (w \curvearrowright u) \in E \setminus E_T \} \cup
     \{ ear_{\ddot{T}_u^{\curvearrowright}}(w) \mid (w \rightarrow u) \in E_{T} \}  \cup
     \{ \overset{\curvearrowright}{\infty} \} )$.

Clearly, the transformation need not be carried out explicitly.
Therefore, we can identified the fictitious vertex $\tilde{u}$ with the corresponding vertex $u$.
The definition of $ear(w)$ can be simplified to:

 $ear(w) =
ear_{\dot{T}_w^{\curvearrowright}}(w)
= \min_{\lessdot} (
     \{ ear_{\ddot{T}_u^{\curvearrowright}}(w) \mid (w \curvearrowright u) \in E \setminus E_T  \vee  (w \rightarrow u) \in E_{T} \}  \cup
     \{ \overset{\curvearrowright}{\infty} \} )$.

\vspace{9pt}

During the $\mathit{dfs}$,
along with $\sigma(w), w \in V$, we maintain an edge set $\alpha(w)$
consisting of all the edges whose end-vertices are in $\sigma(w)$, and
the edges of which one end-vertex is in $\sigma(w)$ while
the other end-vertex is known to be in $\sigma(w)$ but have not been added to $\sigma(w)$.
When  a vertex $w$ is ejected, $\sigma(w)$ is the 3ecc containing $w$, and
$\alpha(w)$ is the edge set of the auxiliary subgraph for $\sigma(w)$.

The key idea of our algorithm is to maintain the following invariant at every vertex during the $\mathit{dfs}$:

``When the adjacency list $L[w]$  is completely processed,
let the $w$-path be:
                 $P_w: w(=w_0) w_1 \ldots w_k (k \ge 0)$.
Then, $\alpha(w)$ consists of the edges of $G_{\langle \sigma(w) \rangle}$ in
$E_{\dot{T}^{\curvearrowright}_{w}} \setminus (E_{\dot{T}^{\curvearrowright}_{w_1}} \cup
   \{(parent_{\dot{T}^{\curvearrowright}_{w}}(w_1) \rightarrow w_1)\})$
(i.e., the edges in $\dot{T}^{\curvearrowright}_{w}$ excluding those in $\dot{T}^{\curvearrowright}_{w_1}$ and the parent edge of $w_1$ in $\dot{T}^{\curvearrowright}_{w}$)  if $k > 0$,
 or
the edges in $E_{\dot{T}^{\curvearrowright}_{w}} \setminus \{ ear(w)\}$
(i.e., the edges in $\dot{T}^{\curvearrowright}_{w}$ excluding $ear(w)$)   if $k=0$.''

  If $w=r$,
 then $k = 0$ and
$\alpha(r) = E_{\dot{T}^{\curvearrowright}_{r}} \setminus \{ ear(r)\}
= E_{\dot{T}^{\curvearrowright}_{r}} \setminus \{ \overset{\curvearrowright}{\infty} \}
= E_{\dot{T}^{\curvearrowright}_{r}}$
  is the edge set of $\acute{G}_{\langle \sigma(r) \rangle}$.
Recall that $\acute{G}$ results from $G$ after the 2-cuts in the latter are replaced by
their corresponding auxiliary edges (see Lemma~\ref{auxiiary-2}).
Hence, $\acute{G}_{\langle \sigma(r) \rangle}$ is the auxiliary 3ecc subgraphs whose edge set is $ \alpha(r) $.
Moreover, the auxiliary subgraphs for the other 3eccs have all been generated.

If $w \neq r$,  after the $\mathit{dfs}$ backtracked to the parent vertex $v$ of $w$,
if $deg_{\hat{G}_w}(w) = 1$, then $k = 0$ and
$\alpha(w) = E_{\dot{T}^{\curvearrowright}_{w}} \setminus \{ ear(w)\}
= E_{\dot{T}^{\curvearrowright}_{w}} \setminus \{ \overset{\curvearrowright}{\infty}\}
= E_{\dot{T}^{\curvearrowright}_{w}}$
  is the edge set of $\acute{G}_{\langle \sigma(w) \rangle}$.
 If $deg_{\hat{G}_w}(w) = 2$,
  by Lemma~\ref{2cut-iff-2},
$\{ (v \rightarrow w), (parent_{\dot{T}^{\curvearrowright}_{w}}(w_1) \rightarrow w_1)\}$
or
$\{(v \rightarrow w), ear(w)\}$
   is the corresponding 2-cut in $\hat{G}_w$.
In the former case,
by the invariant,
   $\alpha(w)$ consists of the edges of $G_{\langle \sigma(w) \rangle}$ in
   $E_{\dot{T}_{w}} \setminus (E_{\dot{T}^{\curvearrowright}_{w_1}} \cup
\{(parent_{\dot{T}^{\curvearrowright}_{w}}(w_1) \rightarrow w_1)\})$.
After adding the edge $(w, parent_{\dot{T}^{\curvearrowright}_{w}}(w_1))$ to $\alpha(w)$,
the resulting $\alpha(w)$ is the edge set of  $\acute{G}_{\langle \sigma(w) \rangle}$.
In the latter case, $k=0$ and
by the invariant,
   $\alpha(w)$ consists of the edges of $G_{\langle \sigma(w) \rangle}$ in
   $E_{\dot{T}^{\curvearrowright}_{w}} \setminus \{ ear(w)\}$.
After adding the edge $(w, s(ear(w))$ to $\alpha(w)$,
 $\alpha(w)$ is the edge set of  $\acute{G}_{\langle \sigma(w) \rangle}$.

\subsection{Generating the auxiliary 3ecc subgraphs}


\vspace{-6pt}



The following is a pseudo-code for the algorithm which is based on the pseudo-code in~\cite{T07}.
During the $\mathit{dfs}$, $\alpha(w), w \in V$, consists of the edges that have been confirmed to be in $\acute{G}_{\langle \sigma(w) \rangle}$.
When $\sigma(w)$ is ejected, $\alpha(w)$ is the edge set of $\acute{G}_{\langle \sigma(w) \rangle}$.
The $w$-path ($u$-path, respective;y)
$\mathcal{P}_w: w (=w_0) w_1 \ldots w_k$, is represented by an array $next( 1.. |V|)$
such that $next(w_i) = w_{i+1}, 0 \le i < k,$ and $next(w_k) = \perp$.
 In \textbf{Procedure} \texttt{construct-aux-subgraph},
the \textbf{for each} loop processes the adjacency list $L[w]$ of vertex $w$.
  The if-part of the first \textbf{if} statement in the \textbf{for} loop deals with
  unvisited vertices leading to child edges while the else-part deals with visited vertices leading to back-edges.
   \textbf{Procedure} \texttt{Gen-aux-edges}
   ejects the supervertex $u$ and adds auxiliary edges to  $\alpha(u)$ based on Theorem~\ref{auxiiary-2} and
Lemma~\ref{2cut-iff-2}
   to create $\acute{G}_{\langle \sigma(w) \rangle}$.
\textbf{Procedure} \texttt{Absorb-path} absorbs the entire $u$-path or $w$-path
  and update $\alpha(w)$ accordingly.
\textbf{Procedure} \texttt{Absorb-subpath} absorbs a section
   of the $w$-path and update $\alpha(w)$ accordingly.
For clarity, instructions for generating the 3eccs are excluded.
%


\begin{singlespacing}

\noindent\hrulefill

\noindent \textbf{Algorithm} \texttt{Auxiliary-3ecc-subgraphs}

\vspace{-6pt}
\noindent\hrulefill

\footnotesize

\noindent \textbf{Input:} A connected multigraph graph $G=(V,E)$ represented by adjacency lists $L[w], \forall w \in V$

\noindent \textbf{Output:}  The auxiliary subgraphs for the 3eccs of $G$:
$\{ \acute{G}_{{\langle \sigma(w) \rangle}} \mid w \in V \}$

\noindent \textbf{begin}

   \textbf{for each} $w \in V$ \textbf{do}     

\hspace{30pt}   $ dfs(w) := 0;$         \hspace{60pt}  // mark $w$ as unvisited

\hspace{30pt}   $ deg(w) := 0;  \
                  nd(w) := 1;$          \hspace{10pt}  // degree of $w$ and number of descendants of $w$

\hspace{30pt}    $parent(w) := \perp$;     

\hspace{30pt}
    $\alpha(w) := \emptyset$;      \hspace{70pt}  // the edge set of $\acute{G}_{{\langle \sigma(w) \rangle}}$

\hspace{30pt}
    $Inc(w) := \emptyset$;      \hspace{60pt}  // a temporary storage for the incoming back-edges

 $cnt := 1$; \ \ \  \hspace{20pt}  // $\mathit{dfs}$ number counter

 \texttt{construct-aux-subgraph}$(r, \perp)$; \ \ \ \ \ // start $\mathit{dfs}$ from vertex $r$

 \textbf{output}$(\alpha(r))$;

\noindent \textbf{end.}\\

%

\noindent \textbf{Procedure} \texttt{construct-aux-subgraph}$(w, v)$

\noindent \textbf{begin}

 $dfs(w) := cnt; \ cnt := cnt + 1; \ parent(w) := v$;  \ \ \
 $  next(w) := \perp$;  \  // $\mathcal{P}_w: w$

$ear( w) := \overset{\curvearrowright}{\infty}$;    \hspace{20pt}  // initialize $ear(w)$

\vspace{3pt}
 \textbf{for each} $u \in L[w]$ \textbf{do}
     \ \ \  \hspace{100pt} // pick the next vertex $u$ in the adjacency list of $w$

\hspace{20pt} $deg(w) := deg(w) + 1$;

 \hspace{20pt} \textbf{if} $(dfs(u) = 0)$ \textbf{then}
  \ \ \ \hspace{7pt} // $u$ is unvisited

\hspace{40pt}
     $Null := false$;
      \ \ \  \hspace{5pt} // $Null = true$ iff $\mathcal{P}_u$ vanishes

  \hspace{40pt} \texttt{construct-aux-subgraph}$(u,w)$;
    \hspace{10pt} // $\mathit{dfs}$ advances to $u$

\hspace{40pt}
     $nd(w) := nd(w) + nd(u)$;   \hspace{55pt}   // update $nd(w)$

\hspace{40pt}
    \textbf{if} $(deg(u) \le 2)$ \textbf{then}      
      \hspace{68pt} // found a 1-cut or 2-cut

     \hspace{60pt}    \texttt{Gen-aux-edges}$( w, u, Null )$;
  \ \  \hspace{3pt} // eject $u$ from $\mathcal{P}_u$ and generate $\acute{G}_{{\langle \sigma(u) \rangle}}$

     \hspace{40pt} \textbf{if} $(ear(w) \ \underline{\lessdot} \ ear(u))$ \textbf{then}
  \hspace{60pt}  // equivalent to $(lowpt(w) \le lowpt(u))$ in~\cite{T07}

\hspace{60pt} \texttt{Absorb-ear($w, u, Null$)}
   \ \ \ \  \hspace{110pt}  // $w$ absorbs the  $u$-path

\hspace{60pt}   \textbf{if} $( Null  \wedge  (t(ear(u)) \prec w) )$
\textbf{then} $\alpha(w) := \alpha(w)  \cup  \{ ear(u) \}$;
 \hspace{5pt}  // $\mathcal{P}_u$ vanishes,  $(w \rightarrow u)$ is not a 1-cut;

 \hspace{4.2in}   // $ear(u)= (w \curvearrowright t(ear(u))$

  \hspace{40pt} \textbf{else}
                \hspace{10pt}  // equivalent to $(lowpt(w) > lowpt(u))$ in~\cite{T07}


           \hspace{70pt}   \texttt{Absorb-ear}$(w, next(w), false)$;
                                         \hspace{52pt}   // $w$ absorbs the current $w$-path

           \hspace{70pt}  \textbf{if} $(Null)$ \textbf{then} $next(w) := \perp$ \textbf{else} $next(w) := u$;
\ \ \ \ \    // attach  $\mathcal{P}_u$ to $w$ to form a new current $\mathcal{P}_w$-path

           \hspace{70pt}
 $ear( w) := ear( u)$;              \hspace{120pt} //    update $ear(w)$

 \hspace{20pt} \textbf{else if} $(dfs(u) < dfs(w) \wedge u \neq parent(w))$ \textbf{then}
   \ \  \hspace{18pt}  // $ w \curvearrowright u$ is an outgoing back-edge of $w$

    \hspace{60pt} \textbf{if} $((w \curvearrowright u) \lessdot ear(w))$ \textbf{then}
      \hspace{60pt}  // equivalent to $(dfs(u) < lowpt(w))$ in~\cite{T07}

       \hspace{75pt}  \texttt{Absorb-ear}$( w, next(w), false )$;      \hspace{32pt} // $w$ absorbs the current $w$-path

        \hspace{75pt}
        $next(w) := \perp$;    \hspace{110pt} //  the new current $w$-path is $\mathcal{P}_w : w$

  \hspace{75pt}             $ear(w) := (w \curvearrowright u)$;
\hspace{85pt} //    update $ear(w)$

    \hspace{62pt} \textbf{else} $\alpha(w) := \alpha(w) \cup \{(w \curvearrowright u)\}$;
          \hspace{50pt}  // absorb $(w \curvearrowright u)$ to update $\alpha(w)$


\hspace{35pt} \textbf{else}  \hspace{160pt}  // an incoming back-edge

 \hspace{50pt}        $Inc(w) := Inc(w) \cup \{ (u \curvearrowright w) \}$;
            \hspace{70pt} // store the incoming back-edge


 \textbf{for each} $(u \curvearrowright w) \in Inc(w)$ \textbf{do}
   \texttt{Absorb-subpath}$( w, u)$;  \hspace{20pt} // dealing with the stored incoming back-edges

\vspace{3pt}
\noindent \textbf{end.} /* of Procedure \texttt{construct-aux-subgraph} */

\vspace{9pt}

\noindent \textbf{Procedure} \texttt{Gen-aux-edges}$( w, u, Null )$  

\noindent \textbf{begin} \ \ \ // $\mathcal{P}_u: (u=)u_0u_1\ldots u_h (h \ge 0)$; \
 eject super-vertex $u$ from $\mathcal{P}_u$ and
                            generate the auxiliary 3ecc subgraph $\acute{G}_{\langle \sigma(u) \rangle}$


 \hspace{-2pt}      \textbf{if} $( deg(u) = 1 )$ \textbf{then} \ \ \
\hspace{70pt} //  $(w,u)$ is a 1-cut

   \hspace{15pt}       $Null  := true$;    \hspace{85pt} //    the $u$-path vanishes after ejecting $u$


 \hspace{-2pt}    \textbf{else} \ \ \ // $(deg_{\hat{G}_u}(u) = 2)$

   \hspace{15pt}
    \textbf{if} ($next(u) = \perp$) \textbf{then}
  \hspace{30pt}  \ \ \   // $\mathcal{P}_u: u$,
i.e., the 2-cut is  $\{(w \rightarrow u), (\ddot{u} \curvearrowright d )\}$,
where $(\ddot{u} \curvearrowright d ) = ear(u)$

\hspace{30pt}
     $\ddot{u} := s(ear(u))$; \ $d := t(ear(u))$;
 \ \ \    \hspace{36pt}
 // determine $\ddot{u}$ and $d$

\hspace{30pt}
     \textbf{if} $( w \neq d)$ \textbf{then} $ear(u) := (w \curvearrowright d)$;
       \ \ \    \hspace{23pt}
 // update $ear(u)$ with the embodiment of $(\ddot{u} \curvearrowright d)$

\hspace{30pt}
     \textbf{if} $(u \neq \ddot{u})$ \textbf{then} $\alpha(u) := \alpha(u) \cup
      \{ (u, \ddot{u}) \}$;
       \ \ \    \hspace{5pt}
 // add auxiliary edge $(u, \ddot{u})$ to   $\acute{G}_{\langle \sigma(u) \rangle}$

\hspace{30pt}
$Null : = true$;     \hspace{112pt}   //   the $u$-path vanishes after ejecting $u$


 \hspace{14pt}
    \textbf{else} $u_1 := next(u); \ \ddot{u} := parent(u_1)$;  \
     \ \ \    \hspace{30pt} //  2-cut is $\{(w \rightarrow u),(\ddot{u} \rightarrow u_1)\}$, where
  $\ddot{u} = parent(u_1)$

\hspace{30pt}
$parent(u_1) := w$;                        
\ \ \  \hspace{92pt}  // replace $\{(w \rightarrow u),(u \rightarrow u_1)\}$ with a new edge $(w \rightarrow u_1)$

\hspace{30pt}
     \textbf{if} $(u \neq \ddot{u})$ \textbf{then} $\alpha(u) := \alpha(u) \cup
      \{ (u, \ddot{u}) \}$;
       \ \ \    \hspace{5pt}
 // add auxiliary edge $(u, \ddot{u})$ to  $\acute{G}_{\langle \sigma(u) \rangle}$

\hspace{30pt}   $u := u_1$;
                                         \ \ \   \hspace{130pt} // remove $u$ from $\mathcal{P}_u$

      \textbf{output}$(\alpha(u))$;
         \hspace{30pt}    //  output the edge set of $\acute{G}_{\langle \sigma(u) \rangle}$

\noindent \textbf{end};


\vspace{9pt}
\noindent \textbf{Procedure} \texttt{Absorb-ear}$( w, u, Null )$

 // absorb path $\mathcal{P}: u u_1 \ldots u_h$, where $\mathcal{P} = \mathcal{P}_u$, or
 $u = next(w)$ on the current $\mathcal{P}_w$ path

\noindent \textbf{begin}

\noindent   \textbf{if} $( not(Null) )$ \textbf{then}
 \hspace{25pt} // rule out the cases where $\mathcal{P}_u$ has been replaced by
                  $(w \curvearrowright t(ear(u))$ or $(w \rightarrow u)$ is a 1-cut

   \textbf{if} $( u = \perp )$ \textbf{then}
 \hspace{90pt} // $\mathcal{P}$ is the current $\mathcal{P}_w$ of the form $\mathcal{P}_w: w$, and

\hspace{15pt}     $\alpha(w) := \alpha(w) \cup \{ ear(w) \}$;
    \hspace{25pt}    // $ear(w) = (w \curvearrowright z)$, where $dsf(z) = lowpt(w)$

   \textbf{else}

\hspace{15pt}    $x := u$;      \ \ \  \hspace{30pt} // absorb $\mathcal{P}$ starting from supervertex $u$

\hspace{15pt}   \textbf{repeat}

\hspace{30pt}     $\alpha(w) := \alpha(w) \cup \alpha(x) \cup \{(parent(x), x)\}$;
 \hspace{23pt}  // absorb supervertex $x$ and the parent edge of  $x$

\hspace{30pt}     $deg(w) := deg(w) + deg(x) - 2$;
 \hspace{58pt}  // update $deg(w)$

\hspace{30pt}     $x := next(x)$;

\hspace{15pt}   \textbf{until}  $ (x = \perp) $;  \ \ \   \hspace{30pt} // the last supervertex has been absorbed

\hspace{15pt}  $\alpha(w) := \alpha(w) \cup \{ ear(u) \}$;
    \hspace{25pt}    // $ear(u) = (u \curvearrowright z)$, where $dsf(z) = lowpt(u)$

\noindent \textbf{end.}

\vspace{12pt}
\noindent \textbf{Procedure} \texttt{Absorb-subpath}$( w, u)$

 // absorb the subpath $\mathcal{P}: w (=w_0) w_1 \ldots w_{\ell} (=u)$ of the current $\mathcal{P}_w$,
 where $next(w_i) = w_{i+1}, 0 \le i < \ell$

\noindent \textbf{begin}

 $deg(w) := deg(w) - 2$;    \hspace{70pt}  // update $deg(w)$

  $x := next(w)$;

     \textbf{while}  $( (x \neq \perp) \wedge (dfs(x) \le dfs(u) < dfs(x) + nd(x)) )$  \textbf{do}
    \ \ \ \ \   // $u$ is a descendant of $x$

\hspace{15pt}     $\alpha(w) := \alpha(w) \cup \alpha(x) \cup \{(parent(x) \rightarrow x)\}$;
 \hspace{23pt}  // absorb supervertex $x$ and the parent edge of  $x$

\hspace{15pt}      $deg(w) := deg(w) + deg(x) - 2$;
 \hspace{70pt}  // update $deg(w)$

\hspace{15pt}       $x := next(x)$;

  $next(w) := x$   \hspace{20pt}  // update the current $\mathcal{P}_w$ by cutting out the subpath $w w_1 \ldots  u$

\noindent \textbf{end}.

\noindent\hrulefill

\normalsize
\end{singlespacing}


\begin{lem}\label{correctness}
   Let $w \in V$.
When the adjacency list $L[w]$ of $w$ is completely processed,

\vspace{-9pt}
\begin{description}
  \item[$(a)$ ] The auxiliary subgraph for each 3ecc residing in the subtrees of $\dot{T}_w^{\curvearrowright}$
    has been constructed;

\vspace{-9pt}
  \item[$(b)$] Let the $w$-path be:
   $\mathcal{P}_w: (w=) w_0 w_1 w_2 \ldots w_k$.
Then,

  $\alpha(w_i) = E_{\dot{T}_{w_i}^{\curvearrowright}} \backslash
  ( E_{\dot{T}_{w_{i+1}}^{\curvearrowright}} \cup \{ (parent_{_{\dot{T}_{w_i}^{\curvearrowright}}}(w_{i+1}) \rightarrow w_{i+1} ) \} ), 0 \le i < k$;

  $\alpha(w_k) = E_{\dot{T}_{w_{k}}^{\curvearrowright}} \setminus
   \{ ear(w) \}, k \ge 0$.

\end{description}

\end{lem}

\vspace{-12pt}
\noindent \textbf{Proof:}
(By induction on the height of $w$ in the $\mathit{dfs}$ tree)

Let $v_1, v_2, \ldots, v_{q} (q = deg_G(w))$, be the vertices in $L[w]$.

 When $w$ is a leaf,
 if $q = 1$, then $w$ is a pendant and Conditions $(a)$ and $(b)$ vacuously hold true.

 Let $q \ge 2$.
 Since  $(w, u_i), 1 \le i \le q,$
 is either an outgoing back-edge or the parent-edge of $w$,
 only the \textbf{then} part of the \textbf{else if} statement within the  \textbf{for each} loop is executed.
 It is easily verified that when $L[w]$ is completely processed,
 $ear(w) =  (z \curvearrowleft w)$ which is the lexicographically smallest outgoing back-edge of $w$, and
 $\alpha(w) = \{ (u_{i} \curvearrowleft w) \mid 1 \le i \le q  \} \backslash \{(z \curvearrowleft w)\}
 = E_{\dot{T}_{w}} \setminus \{ ear(w)  \}$.
Condition $(b)$ thus hold.
 Condition $(a)$ vacuously holds true.

Let $w$ be an internal vertex of $T$.
We shall apply induction to prove that
for $p, 1 \le p \le q,$  after $v_i, 1 \le i \le p$ are processed,
the subgraph of $G$ consisting of
the outgoing back-edges of $w$, $B_p = \{ (w \curvearrowright v_i) \in E \setminus E_T \mid  1 \le i \le p \}$,
 the child edges of $w$, $C_p = \{ ( w \rightarrow v_i) \in E_T \mid 1 \le i \le p \}$ and
 their corresponding  palm trees, $T_p^{\curvearrowright} = \{ T_{v_i}^{\curvearrowright} \mid (w \rightarrow v_i) \in C_p,
 1 \le i \le p \}$
have been transformed into:

\vspace{-12pt}
\begin{description}
  \item[$(a)$] a collection of auxiliary subgraphs each for a distinct 3ecc residing in some palm tree of $T_{p}^{\curvearrowright}$,

\vspace{-9pt}
  \item[$(b)$]   a current $w$-path
 $\mathcal{P}_w: (w=) w_0 w_1 w_2 \ldots w_k (k \ge 0)$ such that:

\vspace{-12pt}
\begin{description}
  \item[$(i)$] $ ear(w) =
\min_{\lessdot} ( \{ ear_{\ddot{T}_{v_i}^{\curvearrowright}}(w) \mid
(w \rightarrow v_i) \in C_p \vee (w \curvearrowright v_i) \in B_p \} \cup
\{ \overset{\curvearrowright}{\infty} \})$;

\vspace{-9pt}
  \item[$(ii)$] $\alpha(w) = \bigcup \{E_{\ddot{T}_{v_i}^{\curvearrowright}}
           \mid ((w \rightarrow v_i) \in C_p \vee  (w \curvearrowright v_i) \in B_p)
\wedge
(ear_{\ddot{T}_{v_i}^{\curvearrowright}}(w) \neq ear(w))  \}$;

\vspace{-9pt}
  \item[$(iii)$]   $\alpha(w_j) =
  E_{\dot{T}_{w_j}^{\curvearrowright}} \backslash (E_{\dot{T}_{w_{j+1}}^{\curvearrowright}} \cup \{ (parent_{\dot{T}_{w_j}^{\curvearrowright}}(w_{j+1}) \rightarrow w_{j+1} ) \}), 1 \le j < k$;

\vspace{-3pt}
  \item[$(iv)$]   $\alpha(w_k) = E_{\dot{T}_{w_{k}}^{\curvearrowright}} \setminus
\{ ear(w) \}, k \ge 1$.

\end{description}
\end{description}

 Consider vertex $v_p$. If $v_p$ is unvisited,
   it becomes a child of $w$.
  When the $\mathit{dfs}$ backtracks from $v_p$ to $w$,

\noindent $(a)$
 $deg_{\hat{G}_{v_p}}(v_p) = 1$:
  Then $(w, v_p)$ is a 1-cut, and $\sigma(v_p)$ is a 3ecc.
 Since $(w, v_p)$ is a 1-cut,
$ear(v_p) = \overset{\curvearrowright}{\infty}$ and
the $v_p$-path is $\mathcal{P}_{v_p}: v_p$.
  By the induction hypothesis on $v_p$,
  Assertion $(b)$ of the lemma holds for $\mathcal{P}_{v_p}$
which implies that
$\alpha(v_p)  =
 E_{\dot{T}_{v_p}^{\curvearrowright}} \setminus
   \{ ear(v_p) \}
= E_{\dot{T}_{v_p}^{\curvearrowright}} \setminus
   \{ \overset{\curvearrowright}{\infty} \}
= E_{\dot{T}_{v_p}^{\curvearrowright}}$.
    Therefore,
         $\alpha(v_p)$ is the edge set of $G_{\langle \sigma(v_p)  \rangle}$ when $v_p$ is ejected.
    As no path connecting two vertices in $G_{\langle \sigma(v_p)  \rangle}$ can use edges outside $G_{\langle \sigma(v_p)  \rangle}$,
         $\forall x, y \in \sigma(v_p), x \overset{3}{\equiv}_{G} y$ if and only if
          $x \overset{3}{\equiv}_{G_{\langle \sigma(v_p)  \rangle}} y$.
         Hence, $G_{\langle \sigma(v_p)  \rangle}$ is a 3-edge-connected subgraph of $G$
     (notice that the induced subgraph is the auxiliary subgraph in this case).

Since $(w, v_p)$ is a 1-cut $\Rightarrow \ddot{T}_{v_p}^{\curvearrowright}$ vanishes,
and $ (w \curvearrowright v_p) \notin B_p$, by the induction hypothesis,

$ear(w) =
\min_{\lessdot} ( \{ ear_{\ddot{T}_{v_i}^{\curvearrowright}}(w) \mid
(w \rightarrow v_i) \in C_{p-1} \vee (w \curvearrowright v_i) \in B_{p-1} \} \cup
\{ \overset{\curvearrowright}{\infty} \})$

\hspace{35pt} $  =
\min_{\lessdot} ( \{ ear_{\ddot{T}_{v_i}^{\curvearrowright}}(w) \mid
 (w \rightarrow v_i) \in C_{p}  \vee (w \curvearrowright v_i) \in B_{p} \}  \cup
\{ \overset{\curvearrowright}{\infty} \})$.
 Condition $(b)(i)$ thus holds.

Since $ear(w)$ remains unchanged,
if $p=1$,
then as $\alpha(w) = E_{\ddot{T}_{v_p}^{\curvearrowright}} = \emptyset$,
 Conditions $(b)(ii)$  follows.
 Conditions $(b)(iii), (iv)$  vacuously hold true.
If $p > 1$,
Conditions $(b)(ii), (iii)$ and $(iv)$ hold
 by the induction hypothesis on $p-1$ and
 $E_{\ddot{T}_{v_p}^{\curvearrowright}} = \emptyset$,
In either case,
  Condition $(a)$ holds by the induction hypothesis applied to $v_p$ and $p-1$, as well as the ejection of $v_p$.

\vspace{9pt}

\noindent $(b)$
 $deg_{\hat{G}_{v_p}}(v_p) \ge 2$:
 if $deg_{\hat{G}_{v_p}}(v_p) = 2$,
 then $\sigma(v_p)$ is a 3ecc.
 The \textbf{else} part of the first \textbf{if} statement in \textbf{Procedure} \texttt{Gen-aux-edges} is executed to eject $v_p$.

\noindent $\bullet$
\ If $next(v_p) \neq \perp$, let $v^1_p = next(v_p)$.
The 2-cut is $\{ (w,v_p), (v_p, v^1_p) \}$.
By Lemma~\ref{2cut-iff-2}$(i)$, the corresponding 2-cut in $\dot{T}_w^{\curvearrowright}$ is
$\{ (w, v_p), (\ddot{v}_p \rightarrow v^1_p)\}$,
where $\ddot{v}_p = parent_{\dot{T}_{v_p}^{\curvearrowright}}(v^1_p)$.
The \textbf{else} part of the second \textbf{if} statement is executed.
Letting $parent(v^1_p) = w$ results in replacing the edges in the 2-cut with a new edge
 $(w, v^1_p)$.
 Moreover,
a new edge $(v_p, \ddot{v}_p)$ is added to $\alpha(v_p)$ if it is not a self-loop.
  Let $G_{\alpha(v_p)}$ be the graph whose edge set is $\alpha(v_p)$.
  By Theorem~\ref{auxiiary-2},
  $G_{\alpha(v_p)}$ is the auxiliary 3ecc subgraph for $\sigma(v_p)$.

\noindent $\bullet$
\
If $next(v_p) = \perp$,
the 2-cut is $\{ (w,v_p), (v_p \curvearrowright d) \}$.
By Lemma~\ref{2cut-iff-2}$(ii)$, the corresponding 2-cut in $\dot{T}_w^{\curvearrowright}$ is
 $\{ (w, v_p), ear(v_p)  \} = \{ (w, v_p), (\ddot{v}_p \curvearrowright d)\}$, where
$\ddot{v}_p = s(ear(v_p))$ and  $d = t(ear(v_p))$.
The \textbf{then} part of the second \textbf{if} statement is executed
resulting in adding
   the edge $(w \curvearrowright d)$  to $\hat{G}_{v_p}$ if it is not a self-loop, and
   the edge $(v_p, \ddot{v}_p)$  to $\alpha(v_p)$ if it is not a self-loop.
  Let $G_{\alpha(v_p)}$ be the graph whose edge set is $\alpha(v_p)$.
By Theorem~\ref{auxiiary-2},
  $G_{\alpha(v_p)}$ is the auxiliary 3ecc subgraph for $\sigma(v_p)$.

\vspace{6pt}
Now,
let the $v_p$-path be $v_p^1 v_p^2 \ldots v_p^h (h \ge 0)$, where
 $v_p^1 = v_p$ if $v_p$ is not ejected, and
 $v_p^1 = next(v_p)$ or vanished, otherwise.
Then,
$\ddot{T}_{v_p}^{\curvearrowright} = (w \rightarrow v_p^1) \cup \dot{T}_{v_p^1}^{\curvearrowright}$
if $h > 0$, or
$\ddot{T}_{v_p}^{\curvearrowright} = (w \curvearrowright t(ear(v_p))$ if $h = 0$.

After $v_p$ is completely processed,
by the induction hypothesis applied to $v_p$ and $p-1$, as well as the ejection of $v_p$ if happened,
 it is easily verified that
Condition $(a)$ holds.
It remains to prove Conditions $(b)(i)$-$(iv)$ hold.

First, consider  $p=1$.                  
Then $v_p = v_1$, $ear(w)= \overset{\curvearrowright}{\infty}$ and $\alpha(w) = \emptyset$.
Since $(w \rightarrow v_1)$ is not a 1-cut, $ear(v_1) \neq \overset{\curvearrowright}{\infty}$
which implies that $ear(v_1) \neq ear(w)$.

If $\underline{ear(v_1) \lessdot ear(w)}$,
the \textbf{else} part of the third \textbf{if} statement within the \textbf{for each} loop is executed
resulting in
$\mathcal{P}_w: w (=w_0) w_1 w_2 \ldots w_h (h \ge 0)$, where $w_i = v_1^i$,
and
$ear(w) =
ear(v_1)$.
%

If $h \ge 1$,
since
$
ear (w)
= ear(v_1)
\Rightarrow
ear (w)
= ear_{\ddot{T}_{v_1}^{\curvearrowright}}(w)$
by Lemma~\ref{palm-tree-ear},
$ear (w) =
\min_{\lessdot} (\{ ear_{\ddot{T}_{v_i}^{\curvearrowright}}(w) \mid (w \rightarrow v_i) \in C_1 )
 \vee (w \curvearrowright v_i) \in B_{1} \} \cup
\{\overset{\curvearrowright}{\infty}\} )$,
            Condition $(b) (i)$ holds.
Moreover, as  $(w \rightarrow v_1) \in C_1$ and $ear_{\ddot{T}_{v_1}^{\curvearrowright}}(w) = ear(w)$,
$\bigcup \{E_{\ddot{T}_{v_i}^{\curvearrowright}}
           \mid ((w \rightarrow v_i) \in C_1 \vee  (w \curvearrowright v_i) \in B_1)
\wedge (ear_{\ddot{T}_{v_i}^{\curvearrowright}}(w) \neq ear(w)) \}
            = \emptyset = \alpha(w)$.
            Condition $(b) (ii)$ follows.
Conditions $(b)(iii)$ and $(iv)$ follows from the induction hypothesis on $v_1$.


If $h = 0$,
the third \textbf{if} statement in \textbf{Procedure} \texttt{Gen-aux-edges}
 sets $ear(v_1) =  (w \curvearrowright d)$,
 where $d = t(ear(v_1))$.
By Lemma~\ref{palm-tree-ear}, $ear_{\ddot{T}_{v_1}^{\curvearrowright}}(w) =   (w \curvearrowright d)$.
We thus have $ear(v_1) = ear_{\ddot{T}_{v_1}^{\curvearrowright}}(w)$.
 After executing  $ear(w) := ear(v_1)$ in the
 \textbf{else} part of the third \textbf{if} statement within the \textbf{for each} loop,
 $ear(w) = ear_{\ddot{T}_{v_1}^{\curvearrowright}}(w)$.
Then,
$ ear (w)
=
\min_{\lessdot} (\{ ear_{\ddot{T}_{v_i}^{\curvearrowright}}(w) \mid (w \rightarrow v_i) \in C_1 )
 \vee (w \curvearrowright v_i) \in B_{1} \} \cup
\{\overset{\curvearrowright}{\infty}\} )$,
            Condition $(b) (i)$ holds.
Moreover,
$\bigcup \{E_{\ddot{T}_{v_i}^{\curvearrowright}}
           \mid ((w \rightarrow v_i) \in C_1 \vee  (w \curvearrowright v_i) \in B_1)
\wedge (ear_{\ddot{T}_{v_i}^{\curvearrowright}}(w) \neq ear(w)) \}
 =  \emptyset
 = \alpha(w)$,
    Condition $(b) (ii)$ follows.
Since $h=0$, Conditions $(b) (iii)$ and $(iv)$ hold true vacuously.



If $\underline{ear(w) \lessdot ear(v_1)}$,
the \textbf{then} part of the third \textbf{if} statement within the \textbf{for each} loop is executed.

If $h \ge 1$,
 vertex $w$ absorbs the $v_1$-path.
 \textbf{Procedure} \texttt{Absorb-path} is  invoked to add the edges
  $\{ (w \rightarrow v_1^1) \} \cup
 \bigcup_{j=1}^{h-1} (\alpha(v_1^j) \cup \{ (parent_{\dot{T}_{v_1^j}^{\curvearrowright}}(v_1^{j+1}) \rightarrow v_1^{j+1})\}) \cup
\alpha(v_1^h) \cup
\{ ear(v_1) \}$
to
$\alpha(w)$.
By the induction hypothesis on $v_1$,
  $\alpha(v_1^j) = E_{\dot{T}_{v_1^j}^{\curvearrowright}} \setminus ( E_{\dot{T}_{v_1^{j+1}}^{\curvearrowright}} \cup
  \{ (parent_{\dot{T}_{v_1^j}}(v_1^{j+1}) \rightarrow v_1^{j+1}) \} ), 1 \le j < h,$  and
  $\alpha(v_1^h) = E_{\dot{T}_{v_1^h}^{\curvearrowright}} \setminus \{ ear(v_1)\}$.
Then,
 $\alpha(w) =
 \{ (w \rightarrow v_1^1) \} \cup
 \bigcup_{j=1}^{h-1} ( E_{\dot{T}_{v_1^j}^{\curvearrowright}} \setminus ( E_{\dot{T}_{v_1^{j+1}}^{\curvearrowright}} \cup
  \{ (parent_{\dot{T}_{v_1^j}^{\curvearrowright}}(v_1^{j+1}) \rightarrow v_1^{j+1}) \} )
   \cup \{ (parent_{\dot{T}_{v_1^{j}}^{\curvearrowright}}(v_1^{j+1}) \rightarrow v_1^{j+1}) \} )
   \cup E_{\dot{T}_{v_1^h}^{\curvearrowright}} \setminus \{ ear(v_1) \}
    \cup \{ ear(v_1) \}$
 $ =  \{ (w \rightarrow v_1^{1}) \}   \cup  E_{\dot{T}_{v_1^1}^{\curvearrowright}}
   =  E_{\ddot{T}_{v_1}^{\curvearrowright}}$.
Since $ear(v_1) \neq ear(w)$,
by Lemma~\ref{palm-tree-ear},
 $ear_{\ddot{T}_{v_1}^{\curvearrowright}}(w) = ear(v_1)
 \Rightarrow
 ear_{\ddot{T}_{v_1}^{\curvearrowright}}(w) \neq ear(w)
 \Rightarrow
 \bigcup \{E_{\ddot{T}_{v_i}^{\curvearrowright}}
           \mid ((w \rightarrow v_i) \in C_1 \vee  (w \curvearrowright v_i) \in B_1)
\wedge (ear_{\ddot{T}_{v_i}^{\curvearrowright}}(w) \neq ear(w)) \}
   =  E_{\ddot{T}_{v_1}^{\curvearrowright}}
   = \alpha(w)$.
 Condition $(b)(ii)$ thus holds.
Moreover, as
$ear(w) \lessdot ear(v_1)
\Rightarrow \overset{\curvearrowright}{\infty}  \lessdot ear(v_1)
\Rightarrow \overset{\curvearrowright}{\infty} \lessdot ear_{\ddot{T}_{v_1}^{\curvearrowright}}(w)$,
$\min_{\lessdot} ( \{ ear_{\ddot{T}_{v_i}^{\curvearrowright}}(w) \mid
(w \rightarrow v_i) \in C_1 \vee (w \curvearrowright v_i) \in B_1 \} \cup
\{ \overset{\curvearrowright}{\infty} \})
= \min_{\lessdot} \{ear_{\ddot{T}_{v_1}^{\curvearrowright}}(w), \overset{\curvearrowright}{\infty}\}
= \overset{\curvearrowright}{\infty}
= ear(w)$.
 Condition $(b)(i)$ thus holds.
Since $\mathcal{P}_w: w$,
Conditions $(b)$ $(iii)$ and $(iv)$ vacuously hold true.


If $h =0$, then $(w \rightarrow v_1)$ is a cut-edge.
 By Lemma~\ref{palm-tree},
$\ddot{T}_{v_1}^{\curvearrowright} = (w \curvearrowright t(ear(v_1)))
\Rightarrow E_{\ddot{T}_{v_1}^{\curvearrowright}}
=\{(w \curvearrowright t(ear(v_1)))\}$.
Then,
$ear(w) =  \overset{\curvearrowright}{\infty} \wedge ear(w) \lessdot ear(v_1)
 \Rightarrow t(ear(v_1)) = w
 \Rightarrow E_{\ddot{T}_{v_1}^{\curvearrowright}}
=\{(w \curvearrowright w) \}$.
Since $(w, w)$ is a self-loop which is discarded,
we thus have $E_{\ddot{T}_{v_1}^{\curvearrowright}} = \emptyset$
 which implies that
$\bigcup \{E_{\ddot{T}_{v_i}^{\curvearrowright}}
           \mid ((w \rightarrow v_i) \in C_1 \vee  (w \curvearrowright v_i) \in B_1)
\wedge (ear_{\ddot{T}_{v_i}^{\curvearrowright}}(w) \neq ear(w)) \}
 = E_{\ddot{T}_{v_1}^{\curvearrowright}} = \emptyset
 = \alpha(w)$.
 Conditions $(b) (ii)$ thus holds.
The proof for Conditions $(b) (i)$ is same as the above case where $h \ge 1$.
   Since $h=0$, Conditions $(b) (iii)$ and $(iv)$ vacuously hold true.

Next, consider $p > 1$.

%
If $\underline{ear(v_p) \lessdot ear(w)}$,
 \textbf{Procedure} \texttt{Absorb-path} is invoked to absorb the current $w$-path.
 Let $v_{\ell}, 1 \le \ell <p$, be such that
 $ear(w) = (w \curvearrowright v_{\ell})$ or
 $ear(w) = ear_{\ddot{T}_{v_{\ell}}^{\curvearrowright}}(w)$.
 Then the current $w$-path is $(w \curvearrowright v_{\ell})$ or
 $w v_{\ell}^1 v_{\ell}^2 \ldots v_{\ell}^k (k \ge 0)$.
 In the former case, $(w \curvearrowright v_{\ell})$ is added  to $\alpha(w)$.
 Then
 as $\ddot{T}_{v_{\ell}}^{\curvearrowright} = (w \curvearrowright v_{\ell})$,
 $E_{\ddot{T}_{v_{\ell}}^{\curvearrowright}} = \{  (w \curvearrowright v_{\ell}) \}$ is added to $\alpha(w)$.
 In the latter case,
 For $k \ge 1$, similar to the case where $p=1$ and $ear(w) \lessdot ear(v_p)$,
 absorbing the path results in adding
the edge set
 $\{ (w \rightarrow v_{\ell}^1) \} \cup \bigcup_{j=1}^{k-1} (\alpha(v_{\ell}^j) \cup \{ (parent_{\dot{T}_{v_{\ell}^j}^{\curvearrowright}}(v_{\ell}^{j+1}) \rightarrow v_{\ell}^{j+1})\}) \cup
\alpha(v_{\ell}^{k}) \cup
\{ ear(v_{\ell}) \}$
which is
 $E_{\ddot{T}_{v_{\ell}}^{\curvearrowright}}$  to $\alpha(w)$.
For $k = 0$,
the current $w$-path is $(w \curvearrowright t(ear(v_{\ell})))$
which is added to $\alpha(w)$.
By Lemma~\ref{palm-tree},
$\ddot{T}_{v_{\ell}}^{\curvearrowright} =
(w \curvearrowright t(ear(v_{\ell})))$
 which implies $E_{\ddot{T}_{v_{\ell}}^{\curvearrowright}}$ is added to $\alpha(w)$.
 Hence, in all cases,  the edge set $E_{\ddot{T}_{v_{\ell}}^{\curvearrowright}}$ is added to $\alpha(w)$.
Before adding $E_{\ddot{T}_{v_{\ell}}^{\curvearrowright}}$ to $\alpha(w)$,
By the induction hypothesis on $p-1$,
 $\alpha(w) = \bigcup \{E_{\ddot{T}_{v_i}^{\curvearrowright}}
           \mid ((w \rightarrow v_i) \in C_{p-1} \vee  (w \curvearrowright v_i) \in B_{p-1})
\wedge (ear_{\ddot{T}_{v_i}^{\curvearrowright}}(w) \neq ear(w)) \}$.
   After executing $ear(w) := ear(v_p) (= ear_{\ddot{T}_{v_p}^{\curvearrowright}}(w))$ and adding $E_{\ddot{T}_{v_{\ell}}^{\curvearrowright}}$ to $\alpha(w)$,
as $ear_{\ddot{T}_{v_p}^{\curvearrowright}}(w) \notin \{ ear_{\ddot{T}_{v_{\ell}}^{\curvearrowright}}(w), (w \curvearrowright v_{\ell}) \}$,
we have
$\alpha(w) = \bigcup \{E_{\ddot{T}_{v_i}^{\curvearrowright}}
           \mid ((w \rightarrow v_i) \in C_{p-1} \vee  (w \curvearrowright v_i) \in B_{p-1})
\wedge (ear_{\ddot{T}_{v_i}^{\curvearrowright}}(w) \neq ear(w)) \}
\cup E_{\ddot{T}_{v_{\ell}}^{\curvearrowright}}
= \bigcup \{E_{\ddot{T}_{v_i}^{\curvearrowright}}
           \mid ((w \rightarrow v_i) \in C_{p} \vee  (w \curvearrowright v_i) \in B_{p})
\wedge (ear_{\ddot{T}_{v_i}^{\curvearrowright}}(w) \neq ear(w)) \}$.
  Condition $(b) (ii)$ thus holds.

 \vspace{6pt}
 The $v_p$-path is then attached to $w$ to form the current $w$-path.


If $h \ge 1$,
 the current $w$-path becomes $\mathcal{P}_w: w (=w_0) w_1 w_2 \ldots w_h$, where
 $w_i = v_p^i, 1 \le i \le h$.
By the induction hypothesis on $p-1$,
$ ear(w) =
\min_{\lessdot} ( \{ ear_{\ddot{T}_{v_i}^{\curvearrowright}}(w) \mid
(w \rightarrow v_i) \in C_{p-1} \vee (w \curvearrowright v_i) \in B_{p-1} \} \cup
\{ \overset{\curvearrowright}{\infty} \})$.
Since $ear(v_p) \lessdot ear(w)  \Rightarrow  ear_{\ddot{T}_{v_p}^{\curvearrowright}}(w) \lessdot ear(w)$,
$ear_{\ddot{T}_{v_p}^{\curvearrowright}}(w) =
\min_{\lessdot} \{ ear(w), ear_{\ddot{T}_{v_p}^{\curvearrowright}}(w) \}
= \min_{\lessdot} ( \{ ear_{\ddot{T}_{v_i}^{\curvearrowright}}(w) \mid
(w \rightarrow v_i) \in C_{p-1} \vee (w \curvearrowright v_i) \in B_{p-1} \} \cup
\{ \overset{\curvearrowright}{\infty} \}
\cup
\{ ear_{\ddot{T}_{v_p}^{\curvearrowright}}(w) \})$
$=
\min_{\lessdot} ( \{ ear_{\ddot{T}_{v_i}^{\curvearrowright}}(w) \mid
(w \rightarrow v_i) \in C_{p} \vee (w \curvearrowright v_i) \in B_{p} \} \cup
\{ \overset{\curvearrowright}{\infty} \})$.
Therefore,
after $ear(w) := ear(v_p) (= ear_{\ddot{T}_{v_p}^{\curvearrowright}}(w))$ is executed,
 Condition $(b)(i)$ holds.
Condition $(b)(iii)$ and $(iv)$ follows from the induction hypothesis on $v_p$.


If $h = 0$,
the third \textbf{if} statement in \textbf{Procedure} \texttt{Gen-aux-edges}
 sets $ear(v_p) =  (w \curvearrowright d)$, where $d = t(ear(v_p))$.
By Lemma~\ref{palm-tree-ear}, $ear_{\ddot{T}_{v_p}^{\curvearrowright}}(w) =   (w \curvearrowright d)$.
We thus have $ear(v_p) = ear_{\ddot{T}_{v_p}^{\curvearrowright}}(w)$.
The remaining argument is same as the above case where $h \ge 1$.  Condition $(b) (i)$ thus holds.
Conditions $(b) (iii)$ and $(iv)$ vacuously hold true.

\vspace{3pt}
If $\underline{ear(w) \lessdot ear(v_p)}$,
 the current $\mathcal{P}_w$-path remains unchanged.
Conditions $(b)$ $(iii)$ and $(iv)$ thus hold by the induction hypothesis on $p-1$.


If $h \ge 1$, $w$ absorbs the $v_p$-path.
 \textbf{Procedure} \texttt{Absorb-path} is thus invoked.
Similar to the  case where $p=1$, it is easily verified that the edge set added  to $\alpha(w)$
is $ E_{\ddot{T}_{v_p}^{\curvearrowright}}$.
By the induction hypothesis on $p-1$,
 $\alpha(w) = \bigcup \{E_{\ddot{T}_{v_i}^{\curvearrowright}}
           \mid ((w \rightarrow v_i) \in C_{p-1} \vee  (w \curvearrowright v_i) \in B_{p-1})
\wedge (ear_{\ddot{T}_{v_i}^{\curvearrowright}}(w) \neq ear(w)) \}$.
 By Lemma~\ref{palm-tree-ear},
$ear_{\ddot{T}_{v_p}^{\curvearrowright}}(w) = ear(v_p)$.
 Then, after adding  $ E_{\ddot{T}_{v_p}^{\curvearrowright}}$ to  $\alpha(w)$,
  $ear(w) \lessdot ear(v_p)  \Rightarrow
ear_{\ddot{T}_{v_p}^{\curvearrowright}}(w) \neq ear(w)
\Rightarrow
\alpha(w) = \bigcup \{E_{\ddot{T}_{v_i}^{\curvearrowright}}
           \mid ((w \rightarrow v_i) \in C_{p-1} \vee  (w \curvearrowright v_i) \in B_{p-1})
\wedge (ear_{\ddot{T}_{v_i}^{\curvearrowright}}(w) \neq ear(w)) \}
\cup E_{\ddot{T}_{v_{p}}^{\curvearrowright}}
= \bigcup \{E_{\ddot{T}_{v_i}^{\curvearrowright}}
           \mid ((w \rightarrow v_i) \in C_{p} \vee  (w \curvearrowright v_i) \in B_{p})
\wedge (ear_{\ddot{T}_{v_i}^{\curvearrowright}}(w) \neq ear(w)) \}$.
  Condition $(b) (ii)$ thus holds.
 Moreover,
 as $ear(w) \lessdot ear(v_p)  \Rightarrow  ear(w) \lessdot ear_{\ddot{T}_{v_p}^{\curvearrowright}}(w)$,
   by the induction hypothesis on $p-1$,

$ ear(w) =
\min_{\lessdot} ( \{ ear_{\ddot{T}_{v_i}^{\curvearrowright}}(w) \mid
(w \rightarrow v_i) \in C_{p-1} \vee (w \curvearrowright v_i) \in B_{p-1} \} \cup
\{ \overset{\curvearrowright}{\infty} \})$

\noindent $\Rightarrow ear(w) =
\min_{\lessdot} (( \{ ear_{\ddot{T}_{v_i}^{\curvearrowright}}(w) \mid
(w \rightarrow v_i) \in C_{p-1} \vee (w \curvearrowright v_i) \in B_{p-1} \} \cup
\{ \overset{\curvearrowright}{\infty} \})  \cup
  \{ ear_{\ddot{T}_{v_p}^{\curvearrowright}}(w) \})
   =  \min_{\lessdot} ( \{ ear_{\ddot{T}_{v_i}^{\curvearrowright}}(w) \mid
(w \rightarrow v_i) \in C_{p} \vee (w \curvearrowright v_i) \in B_{p} \} \cup
\{ \overset{\curvearrowright}{\infty} \})$.
   Condition $(b)(i)$ thus holds.


If $h = 0$,
the third \textbf{if} statement in \textbf{Procedure} \texttt{Gen-aux-edges}
 sets $ear(v_p) =  (w \curvearrowright d)$, where
  $d = t(ear(v_p))$.
Then, the fourth \textbf{if} statement within the \textbf{for each} loop adds
$(w \curvearrowright d)$ to $\alpha(w)$.
  Since $h=0 \Rightarrow (w \rightarrow v_p)$ is a cut-edge,
  by Lemma~\ref{palm-tree},
  $\ddot{T}_{v_p}^{\curvearrowright} = (w \curvearrowright d)$.
Hence, the \textbf{if} statement adds
$E_{\ddot{T}_{v_p}^{\curvearrowright}}$ to $\alpha(w)$. The remaining part of the proof for Condition $(b)(ii)$ is same as the above case, where $h \ge 1$.
 The proof for Condition $(b)(i)$ is also same as the above case, where $h \ge 1$.

\vspace{3pt}
If $\underline{ear(w) = ear(v_p)}$, then $ear(w) = ear(v_p) =  \overset{\curvearrowright}{\infty}$
 which implies that
$(w, v_p)$ is a 1-cut.
This reduces to the case where
 $deg_{\hat{G}_{v_p}}(v_p) = 1$.

\vspace{6pt}
If $v_p$ is visited, then if $v_p = parent(w)$, the vertex is skipped;
 if $dfs(v_p) > dfs(w)$, $(v_p \curvearrowright w)$ is an incoming back-edge of $w$ which is stored in $Inc(w)$.
 Otherwise, $dfs(v_p) < dfs(w)$, and
 $(w \curvearrowright v_p)$ is an outgoing back-edge of $w$.

If $p=1$, then as $ear(w) = \overset{\curvearrowright}{\infty}$,
  $(w \curvearrowright v_1) \lessdot ear(w)$.
The \textbf{then}  part of the fifth \textbf{if} statement within the \textbf{for each} loop is executed
resulting in
$ear(w) = (w \curvearrowright v_1) = ear_{\ddot{T}_{v_1}^{\curvearrowright}}(w) =
\min_{\lessdot} ( \{ ear_{\ddot{T}_{v_i}^{\curvearrowright}}(w) \mid
(w \rightarrow v_i) \in C_{1} \vee (w \curvearrowright v_i) \in B_{1} \} \cup
\{ \overset{\curvearrowright}{\infty} \})$.
Condition $(b)(i)$ thus holds.
Since $\alpha(w) = \emptyset$,
 $\bigcup \{E_{\ddot{T}_{v_i}^{\curvearrowright}}
           \mid ((w \rightarrow v_i) \in C_{1} \vee  (w \curvearrowright v_i) \in B_{1})
\wedge (ear_{\ddot{T}_{v_i}^{\curvearrowright}}(w) \neq ear(w)) \} = \emptyset$
  as
$ear_{\ddot{T}_{v_1}^{\curvearrowright}}(w) = ear(w)$.
Condition $(b)(ii)$ thus holds.
Conditions $(b)(iii)$ and $(iv)$ hold vacuously.

For $p > 1$,                               
if  $(w \curvearrowright v_p) \lessdot ear(w)$,
then $w$ absorbs the current $\mathcal{P}_w: w w_1 \ldots w_k$ resulting in $\mathcal{P}_w: w$.
Hence,
Conditions $(b)(iii)$ and $(iv)$ vacuously hold.
The proofs for $(b) (i)$ and $(ii)$ are similar to the above case where $ear(v_p) \lessdot ear(w)$,
$p > 1$ and $ h = 0$.
%
If $ear(w) \lessdot (w \curvearrowright v_p)$,
then as the current $w$-path remains unchanged,
Condition $(b) (iii)$ and $(iv)$ continue to hold.
The proofs for $(b) (i)$ and $(ii)$ are similar to those of the above case where
 $ear(w) \lessdot ear(v_p)$, $p > 1$ and $ h = 0$.

\vspace{6pt}
On exiting the \textbf{for each} loop, $p = q$.
Let the current $w$-path be $\mathcal{P}_w: (w=) w_0 w_1 w_2 \ldots w_k (k \ge 0)$.
By Condition $(b) (ii)$,
$\alpha(w) = \bigcup \{E_{\ddot{T}_{v_i}^{\curvearrowright}}
           \mid ((w \rightarrow v_i) \in C_q \vee  (w \curvearrowright v_i) \in B_q)
\wedge (ear_{\ddot{T}_{v_i}^{\curvearrowright}}(w) \neq ear(w)) \}$

\vspace{6pt}
   $\Rightarrow \alpha(w_0)
 =  \left\{
            \begin{array}{ll}
               E_{\dot{T}_{w_0}^{\curvearrowright}} \setminus ( E_{\dot{T}_{w_1}^{\curvearrowright}} \cup
                \{ ( parent_{\dot{T}_{w_0}^{\curvearrowright}}(w_1) \rightarrow w_1 ) \} ),
                    & \hbox{ if $k > 0$;} \\
               E_{\dot{T}_{w_0}^{\curvearrowright}} \setminus \{ ear(w) \},
                    & \hbox{ if $k = 0$.}
            \end{array}
    \right.$

\vspace{6pt}
   Combining with Conditions $(b) (iii)$ and $(iv)$, we have:

\vspace{6pt}
$
 \left\{
   \begin{array}{ll}
    \alpha(w_j) = E_{\dot{T}_{w_j}^{\curvearrowright}} \backslash (E_{\dot{T}_{w_{j+1}}^{\curvearrowright}} \cup
\{ (parent_{\dot{T}_{w_j}^{\curvearrowright}}(w_{j+1}) \rightarrow w_{j+1} ) \}),   & \hbox{$0 \le j < k$;} \\

     \alpha(w_k) = E_{\dot{T}_{w_{k}}^{\curvearrowright}} \setminus \{ ear(w) \}, & \hbox{$ k \ge 0$.}
   \end{array}
 \right.
$
\vspace{6pt}

Then, the following \textbf{for each} statement takes each edge $(u \curvearrowright w)$ in $Inc(w)$
and call
\textbf{Procedure} \texttt{Absorb-subpath}$(w,u)$.
 It is easily verified that when all edges in $Inc(w)$ are processed,
 the sub-path $w w_1 \ldots w_{\ell}$ on the $w$-path has been absorbed by $w$,
 where
  $\ell (> 0)$ is the largest index for which there exists an $(u \curvearrowright w) \in Inc(w)$
  with $w_{\ell} \preceq u$.
The edges on the sub-path
 $ \bigcup_{i=1}^{\ell} (\alpha(w_i) \cup \{ (parent_{\dot{T}_{w_{i-1}}^{\curvearrowright}}(w_i) \rightarrow w_i)\})$
 are added to $\alpha(w)$.

Then,
 $\alpha(w) = \alpha(w_0) \cup    \bigcup_{i=1}^{\ell} (\alpha(w_i) \cup \{ (parent_{\dot{T}_{w_{i-1}}^{\curvearrowright}}(w_i) \rightarrow w_i)\})$

$= \bigcup_{i=0}^{\ell} \alpha(w_i) \cup
\bigcup_{i=1}^{\ell} \{ (parent_{\dot{T}_{w_{i-1}}^{\curvearrowright}}(w_i) \rightarrow w_i)\}$

$ =  \bigcup_{i=0}^{\ell} ( E_{\dot{T}_{w_i}^{\curvearrowright}} \backslash
(E_{\dot{T}_{w_{i+1}}^{\curvearrowright}} \cup
\{ (parent_{\dot{T}_{w_i}^{\curvearrowright}}(w_{i+1}) \rightarrow w_{i+1} ) \}) ) \cup
\bigcup_{i=1}^{\ell} \{ (parent_{\dot{T}_{w_{i-1}}^{\curvearrowright}}(w_i) \rightarrow w_i)\}$

$\Rightarrow
\alpha(w_0) = \left\{
     \begin{array}{ll}
       E_{\dot{T}_{w_0}^{\curvearrowright}} \setminus
(E_{\dot{T}_{w_{\ell+1}}^{\curvearrowright}} \cup
\{ (parent_{\dot{T}_{w_{\ell}}^{\curvearrowright}}(w_{\ell+1}) \rightarrow w_{\ell+1} ) \}), & \hbox{if $\ell < k$;} \\
       E_{\dot{T}_{w_0}^{\curvearrowright}} \setminus \{ ear(w) \}, & \hbox{if $\ell = k$.}
     \end{array}
   \right.
$

\vspace{6pt}
When $Inc(w)$  is  processed,
the $w$-path becomes $\mathcal{P}_w: w(=w_0) w_{\ell + 1} \ldots w_k$, where

\vspace{6pt}
$\left\{
   \begin{array}{ll}
    \alpha(w_0) = \left\{
     \begin{array}{ll}
       E_{\dot{T}_{w_0}^{\curvearrowright}} \setminus
(E_{\dot{T}_{w_{\ell+1}}^{\curvearrowright}} \cup
\{ (parent_{\dot{T}_{w_{\ell}}^{\curvearrowright}}(w_{\ell+1}) \rightarrow w_{\ell+1} ) \}),
& \hbox{if $\ell < k$,} \\
       E_{\dot{T}_{w_0}^{\curvearrowright}} \setminus \{ ear(w) \},
& \hbox{if $\ell = k$.}
     \end{array}
   \right.;
& \hbox{} \\
     \alpha(w_j) =
  E_{\dot{T}_{w_j}^{\curvearrowright}} \backslash (E_{\dot{T}_{w_{j+1}}^{\curvearrowright}} \cup
\{ (parent_{\dot{T}_{w_j}^{\curvearrowright}}(w_{j+1}) \rightarrow w_{j+1} ) \}),
& \hbox{if $\ell + 1 \le j < k$;} \\
     \alpha(w_k) = E_{\dot{T}_{w_{k}}^{\curvearrowright}} \setminus \{ ear(w) \},
& \hbox{if $ \ell < k$.}
   \end{array}
 \right.
$

\vspace{6pt}
Renumbering $\ell+1, \ell+2 \ldots k$ as $1, 2, \ldots, k'$, where $ k' = k - \ell$,
the $w$-path becomes $\mathcal{P}_w: w(=w_0) w_{1} w_{2} \ldots w_{k'}$, and

$\alpha(w_j) =
  E_{\dot{T}_{w_j}^{\curvearrowright}} \backslash (E_{\dot{T}_{w_{j+1}}^{\curvearrowright}} \cup
\{ (parent_{\dot{T}_{w_j}^{\curvearrowright}}(w_{j+1}) \rightarrow w_{j+1} ) \}), 0 \le j < k'$;

 $\alpha(w_{k'}) = E_{\dot{T}_{w_{k'}}^{\curvearrowright}} \setminus \{ ear(w) \}, k' \ge 0$.
The lemma follows.
   \ \ \ \ \  $\blacksquare$

Notice that, for the sake of clarity, the above proof does not consider parallel edges. To include parallel edges, we can modify Condition $(b)(ii)$ as follows:

 $\alpha(w) = \bigcup \{E_{\ddot{T}_{v_i}^{\curvearrowright}}
           \mid ((w \rightarrow v_i) \in C_p \vee  (w \curvearrowright v_i) \in B_p)
\wedge
( (ear_{\ddot{T}_{v_i}^{\curvearrowright}}(w) \neq ear(w))
\vee (\exists j, 1 \le j < i)(ear_{\ddot{T}_{v_j}^{\curvearrowright}}(w) = ear(w) ) \}$.


\begin{thm}\label{correctness1}
\textbf{Algorithm} Auxiliary-subgraphs constructs an auxiliary subgraph  for every 3-edge-connected component of $G$.
\end{thm}

\noindent \textbf{Proof:}
 The algorithm terminates execution when the adjacency list $L[r]$ of the root $r$ is completely processed,
Since $ear(r) = \overset{\curvearrowright}{\infty}$, the $r$-path is $\mathcal{P}_r: r$, which implies that $k=0$.
By Lemma~\ref{correctness}$(b)$,
$\alpha(r)
= E_{\dot{T}_{r}^{\curvearrowright}} \setminus
   \{ ear(r) \}
= E_{\dot{T}_{r}^{\curvearrowright}} \setminus
   \{\overset{\curvearrowright}{\infty}\}
= E_{\dot{T}_{r}^{\curvearrowright}}$.
Hence, $\alpha(r)$ is the edge set  of $\acute{G}_{\langle \sigma(r) \rangle}$.
Then by Lemma~\ref{correctness}$(a)$, the auxiliary subgraphs for the 3eccs of $G$ are constructed.
 \ \ \ \ \ \  $\blacksquare$


\begin{thm}\label{time}
\textbf{Algorithm} \texttt{Auxiliary-3ecc-subgraphs}  takes $O(|V|+|E|)$ time.
\end{thm}

\vspace{-6pt}
\noindent \textbf{Proof:}
   \textbf{Algorithm} \texttt{Auxiliary-3ecc-subgraphs}  is an extension of
\textbf{Algorithm} \texttt{3-edge-} \texttt{connectivity} of~\cite{T07}.
  The extension includes new instructions for generating $\alpha(w), ear(w),  Inc_w,$ $w \in V$.
   Since \textbf{Algorithm} \texttt{3-edge-}\texttt{connectivity} takes $O(|V|+|E|)$ time~\cite{T07},
it suffices to show that the extension takes $O(|V|+|E|)$ time.

The initialization instructions in the main program takes $O(|V|)$ time.
In \textbf{Procedure} \texttt{construct-aux-subgraph}
(whose counterpart is \textbf{Procedure} \texttt{3-edge-connect} in~\cite{T07}),
 the new instructions for initialization  takes $O(1)$ time.
Each call of \textbf{Procedure} \texttt{Gen-aux-edges} takes $O(1)$ time.
Since there are at most $n$ such calls,
the procedure takes a total of $O(|V|)$ time.
Since computing $ear(w)$ takes $O(deg_G(w))$ time, computing $ear(w), \forall w \in V$,
thus takes $\sum_{v \in V} O(deg_G(w)) = O(|E|)$ time.
The instructions for generating $Inc_w, w \in V,$ takes $O(\sum_{w \in V}|Inc_w|)= O(|E \setminus E_T|)
= O(|E|)$ time. 
When all the edges in $Inc(w)$ are processed by the second \textbf{for each} statement,
let $w w_1 \ldots w_{\ell}$ be the section on the $w$-path absorbed by $w$,
 where
  $\ell (> 0)$ is the largest index for which there exists an $(w \curvearrowleft u) \in Inc(w)$
  with $w_{\ell} \preceq u$.
Since every edge on the section is absorbed once,
the total time spent on calling  \textbf{Procedure} \texttt{Absorb-path} to absorb the section
takes $O(\ell + |Inc_w|)$ time.
Hence, in total, the procedure takes $\sum_{w \in V} O(\ell + |Inc_w|)  = O(|E_T|) + O(|E|) = O(|E|)$ time.
\textbf{Procedure} \texttt{Absorb-ear} takes a total of $O(|E|)$ time.
Since every edge is added to an $\alpha(w)$ in $O(1)$ time, and computing $\alpha(w) \cup \alpha(u)$ takes $O(1)$
 as linked lists are used to represent the sets,
  computing $\alpha(w), \forall w \in V$, thus takes $O(|V| + |E|)$.
Computing $deg(v), \forall v \in V$, takes $O(|E|)$ time~\cite{NT14}.
 Hence, the new instructions take $O(|V|+|E|)$ time.
 \textbf{Algorithm} \texttt{Auxiliary-3ecc-subgraphs} thus runs in $O(|V|+|E|)$ time.
\ \ \ \ \ $\blacksquare$

\section{Conclusion}

We have presented a linear-time algorithm for constructing the 3-edge-connected auxiliary subgraphs of a connected undirected multigraph. This algorithm is notably simpler and easier to implement than the previously known best algorithms. Not only does it serve as a valuable preprocessing step for input graphs in 4-edge-connected component algorithms, but it is also applicable to any study involving 3-edge-connected components within a graph.

\vspace{-6pt}

\begin{singlespacing}
\begin{small}

\end{small}

\end{singlespacing}

\end{document}